\documentclass[iop,apj,numberedappendix,twocolappendix,appendixfloats]{emulateapj}
\slugcomment{{}}

\usepackage{epsfig}
\usepackage{amssymb}
\usepackage{amsmath}
\usepackage{graphicx}
\usepackage{natbib}
\usepackage{setspace}

\def \Ledd{$L_{\rm{Edd}}$}
\def \msyr{~$M_{\odot}~\rm{yr}^{-1}$}

\usepackage{mathrsfs}
\usepackage[english]{babel}

\newcommand{\f}[2]{\frac{#1}{#2}} 
\newcommand{\beq}{\begin{equation}}
\newcommand{\seq}{\end{equation}}
\newcommand{\pd}[2]{\frac{\partial #1}{\partial #2}}

\newcommand{\Lnuavg}{\langle \mathscr{L}_\nu\rangle}

\newcommand{\Favg}{\langle F_X \rangle}
\newcommand{\javg}{\langle j_\nu \rangle}
\newcommand{\dL}{\Delta \mathscr{L}_\nu}
\newcommand{\dC}{\Delta L_X(t - \tau)}

\newcommand{\IRF}{\Psi(\nu,t)}
\newcommand{\Ae}{{\rm A}_{\rm Z}}

\shorttitle{Reverberation mapping calculations using a hydrodynamical disk wind solution}
\shortauthors{Waters, Kashi, Proga, Eracleous, Barth, \& Greene}

\begin{document}

\title{
Reverberation mapping of the Broad Line Region: \\
application to a hydrodynamical line-driven disk wind solution
}

\author{
Tim Waters\altaffilmark{1,$\dagger$},
Amit Kashi\altaffilmark{2},
Daniel Proga\altaffilmark{1},
Michael Eracleous\altaffilmark{3,4},
Aaron J. Barth\altaffilmark{5},
and Jenny Greene\altaffilmark{6}
}
\affil{
\altaffilmark{1}{Department of Physics \& Astronomy, University of Nevada, Las Vegas, 4505 S. Maryland Pkwy, Las Vegas, NV, 89154-4002, USA}\\
\altaffilmark{2}{Minnesota Institute for Astrophysics, University of Minnesota, 116 Church St. SE. Minneapolis, MN 55455, USA}\\
\altaffilmark{3}{Department of Astronomy \& Astrophysics and Institute for Gravitation and the Cosmos, \\
The Pennsylvania State University, 525 Davey Lab, University Park, PA~16802, USA}\\
\altaffilmark{4}{Department of Astronomy, University of Washington, Box 351580, Seattle, WA 98195, USA}\\
\altaffilmark{5}{Department of Physics and Astronomy, University of California, Irvine, Irvine, CA 92697, USA}\\
\altaffilmark{6}{Department of Astrophysical Sciences, Princeton University, Princeton, NJ 08544, USA}
}

\altaffiltext{$\dagger$}{email: \href{mailto:waterst3@unlv.nevada.edu}{waterst3@unlv.nevada.edu}}

\begin{abstract}
The latest analysis efforts in reverberation mapping are beginning to allow reconstruction 
of echo images (or velocity-delay maps) that encode information about the structure and 
kinematics of the broad line region (BLR) in active galactic nuclei (AGNs).
Such maps can constrain sophisticated physical models for the BLR.  
The physical picture of the BLR is often theorized to be a photoionized wind launched 
from the AGN accretion disk.  Previously we showed that the line-driven disk wind solution 
found in an earlier simulation by Proga and Kallman is virialized over a large distance 
from the disk.  This finding implies that, according to this model, black hole masses can 
be reliably estimated through reverberation mapping techniques.  However, 
predictions of echo images expected from line-driven disk winds are not available.  
Here, after presenting the necessary radiative transfer methodology, 
we carry out the first calculations of such predictions.  We find that the echo images 
are quite similar to other virialized BLR models such as randomly orbiting clouds and 
thin Keplerian disks.  We conduct a parameter survey exploring how echo images, 
line profiles, and transfer functions depend on both the inclination angle and the line 
opacity.  We find that the line profiles are almost always single peaked, 
while transfer functions tend to have tails extending to large time delays.
The outflow, despite being primarily equatorially directed, 
causes an appreciable blue-shifted excess on both the echo image and line profile 
when seen from lower inclinations ($i \lesssim 45^\circ$).  
This effect may be observable in low ionization lines such as $\rm{H}\beta$.  
\end{abstract}



\section{Introduction}
\label{sec:intro}
\setcounter{footnote}{0}

Broad emission lines have for decades been used as a
basis for classifying active galactic nuclei (AGNs), 
yet the structure and dynamics of the 
broad line region (BLR) around AGNs remains elusive.  
As it will be impossible for the foreseeable future to resolve a BLR 
via direct imaging, we are left with only indirect methods to probe
its spatial and kinematic properties.    
Temporal monitoring observations can
be used to obtain such information using the technique of reverberation mapping (e.g.,
\citealt{BlandfordMcKee1982}; 
\citealt{Peterson1993};
\citealt{Ulrich1997}; 
\citealt{PetersonWandel1999}, \citeyear{PetersonWandel2000};
\citealt{Kaspi2000}; 
\citealt{Krolik2001}; 
\citealt{Peterson2001}, \citeyear{Peterson2006}, \citeyear{Peterson2013}; 
\citealt{Uttley2014}).  

Assuming that the BLR is virialized,
reverberation mapping can be used to estimate the mass of 
the central supermassive black hole (SMBH), $M_{BH}$.  
A measure of the time delay, 
$\langle \tau \rangle$,
for gas to respond to changes in the continuum
determines a characteristic BLR radius $R = c\,\langle \tau \rangle$ (where $c$ is the speed of light),
while the velocity widths of broad emission line profiles
are used to assign a characteristic velocity $\Delta V$.  
The actual black hole mass measurement,
\begin{equation}
M_{BH}=f\frac{R(\Delta V)^2}{G},
\label{eq:f}
\end{equation}
has a potentially major uncertainty associated with the value of $f$, 
the so-called virial coefficient that depends on the geometry and 
kinematics of the BLR.  Furthermore, there can be significant uncertainties associated with 
the measurements of $\langle \tau \rangle$ and $\Delta V$ (e.g.,
Krolik 2001), especially if $\langle \tau \rangle$ is determined by first assuming a form 
for the transfer function (the approach taken in the code \textsc{Javelin}, for example; Zu et al. 2011).
Hence, even for this least demanding application of reverberation mapping, 
it is necessary to look to physical models of the BLR 
that obey observational constraints to better quantify the uncertainties
associated with these quantities.
Several models have been suggested, including randomly orbiting clouds, inflowing and
outflowing gas, rotating disks with thermal or line driven winds, and
more (see, for example, the review by \citealt{MathewsCapriotti1985}
and a more recent summary in Section 5 of \citealt{Sulentic2000}).
 
Although a great deal of work has been done to model the photoionization 
of the BLR gas, relatively few calculations aimed at deriving line profiles and 
transfer functions have been performed, especially ones taking into account 
both hydrodynamics and radiative transfer (e.g., Chiang \& Murray 1996).  
Indeed, the majority of these 
modeling efforts employ stochastic methods (e.g., \citealt{Pancoast2011})
that, while sophisticated,\footnote{
We refer specifically to discrete particle, Monte-Carlo based methods that
model the BLR by prescribing probability distributions for the particles' 
emission properties and kinematics.}
cannot easily incorporate the extensive modeling capability offered by 
performing calculations from first principles using numerical simulations.
In this work, we therefore adopt the complementary approach of calculating
echo images, line profiles and transfer functions by post-processing grid-based 
hydrodynamical simulation data. 

Acceptable theoretical models for the BLR must be able to reproduce the
profiles and relative strengths of the broad emission lines, as well as
their variability properties in response to fluctuations of the ionizing continuum 
on a variety of time scales.  One of the suggested models is a disk wind 
(e.g., \citealt{Shields1977}; \citealt{Emmering1992};
\citealt{ChiangMurray1996}; \citealt{Bottorff1997}), and 
line driving (\citealt{CAK1975}) is one of the common mechanisms by 
which astrophysical objects can launch winds.   
While line driving has been invoked to explain AGN winds (e.g.,
\citealt{Murray1995}; \citealt{Proga2000}), there is no consensus 
that it is the dominant mechanism, as the wind may be
over-ionized by X-ray radiation coming from the central engine,
and in that case the efficiency of line driving is low.  However, 
Proga et al. (2000) showed that clumps forming in the vicinity of the 
SMBH can shield the other parts of the wind from the radiation, 
and enable line-driven winds (see also Proga \& Kallman 2004, hereafter PK04).  

Attributing the BLR to an accretion disk wind is
appealing because this type of model simultaneously provides a framework 
for understanding quasar broad absorption lines (BALs).  Moreover, it does not
require the existence of dense and highly supersonic clouds surrounding the central engine.
Such clouds were shown early on to be prone to rapid destruction due to 
hydrodynamical instabilities (e.g., Mathews 1986; Krolik 1988), 
a finding supported by detailed numerical simulations (Proga et. al 2014; Proga \& Waters 2015).
Several previous investigations suggest that at least part of the
observed line emission originates in a virialized flow, such as a Keplerian disk or a
rotating outflow (e.g., \citealt{Kollatschny2003};
\citealt{CrenshawKraemer2007}; \citealt{Bentz2010b};
\citealt{Kollatschny2013b}; \citealt{Pancoast2014}).
In view of the promise that this family
of models have shown so far, we have embarked on a more extensive
investigation of their observational consequences.
 
For any BLR model to permit the use of equation (1), the responding
gas must be virialized.  Hence, in the case of disk winds, the outflow itself
must be virialized.  A rigorous approach to testing this requirement was
taken by \cite{Kashi2013}, who analyzed various outflow solutions and 
found that the \emph{line-driven wind solution}
presented by PK04 is indeed virialized out to large distances, owing to the
dominance of the rotational component of the wind velocity.  Formally,
a system is virialized if the sum of the density-weighted, volume-integrated 
internal energy and kinetic energy is equal to -1/2 the value of the density-weighted, 
volume-integrated gravitational potential energy (see eqns. 2-3 in Kashi et al. 2013).
Importantly, \cite{Kashi2013} found that the outflow in the PK04 solution
will be observed as virialized from any line of sight (LoS).

In this paper, we extend the investigation of the PK04 solution. It is not
enough to show that the wind is virialized; 
we must quantify how gas responds to variations in the ionizing continuum.
We therefore calculate the observables obtainable from reverberation 
mapping campaigns (namely echo images, emission line profiles, and transfer functions).
Aside from qualitatively understanding how echo images of line-driven disk 
wind solutions differ from the classic examples, it is important to quantify
how the line profiles and transfer functions, as well as the echo images,
depend on optical depth, inclination angle, and kinematics.
The main goal of this paper is to uncover this dependence after presenting
the radiative transfer methodology necessary to perform reverberation mapping
calculations using hydrodynamical disk wind solutions. 

To this end, we adopt very simple, 
parametric prescriptions for the source function in order 
to compare our results with past investigations.   
In forthcoming papers, we will carry out
detailed, self-consistent calculations of the photoionization structure of the wind in order 
to obtain the source function throughout its volume and the dependence of the 
source function on the flux of the ionizing continuum.  Thus, we will be able to
more realistically assess the short-term variability of the broad
emission lines in response to a fluctuating ionizing flux from the
central engine and produce suites of synthetic line profiles meant to
represent populations of AGNs. We defer a quantitative comparison
of the model predictions to the observations to these future papers.

This paper is structured as follows.  In \S{\ref{sec:formalism}},
we present our formalism to derive the impulse 
response function\footnote{
What we call the impulse response function is normally 
termed the 2-D transfer function, an echo image is its digital representation,
and we reserve \emph{transfer function} to explicitly denote the 
frequency-integrated impulse response function.}, 
the fundamental quantity in reverberation mapping.
In \S{\ref{sec:methods}}, we discuss the methods used to evaluate it.  
We apply our methods to the PK04 solution
in \S{\ref{sec:results}}.  We summarize and discuss our results in
\S{\ref{sec:summary}}, and we conclude with a mention of the limitations
of this work and our opinion on how to make further progress in \S{\ref{sec:conclusions}}.

\section{Formalism}
\label{sec:formalism}
The classic work of Blandford \& McKee (1982; hereafter BM82) was published 
a year before the appearance of a seminal paper by 
Rybicki \& Hummer (1983; hereafter RH83), who presented the methodology that 
is now widely used to calculate line profiles in rapidly moving media.  
Therefore, we first derive the impulse response function using the 
framework of RH83, showing how it is consistent with the one first derived by BM82.  

\subsection{Derivation of the Impulse Response Function}
From RH83, the specific monochromatic luminosity $\mathscr{L}_\nu$
due to line emission can be calculated by integrating the 
product of the monochromatic emission coefficient (or emissivity) $j_\nu$ 
and the directional escape probability $\beta_\nu$ 
over the volume $V$ of the entire emitting region:
\beq \mathscr{L}_\nu(t)  = \int dV\, j_\nu(\mathbf{r},t)\beta_\nu(\mathbf{r},t). 
\label{RHdef} \seq
Here, both $j_\nu$ and $\beta_\nu$ depend on the direction of emission, $\hat{n}$; 
only one direction, that pointing toward the distant observer,
contributes to $\mathscr{L}_\nu(t)$.   
The product $j_\nu \beta_\nu$ can be considered an effective
emissivity, the role of $\beta_\nu$ being to allow a unified treatment of optically 
thick and thin gas.  In particular, as demonstrated by 
Chiang \& Murray (1996), 
the escape probability formalism permits a straight forward calculation of 
how optically thick regions \emph{in rapidly moving media} respond to variations 
in the ionizing continuum (through the effects of velocity shear).  In contrast, 
the response from optically thick regions in static or slowly moving media is much 
more difficult to calculate on account of the extra time delays associated 
with multiple scatterings.  
 
To proceed, a distinction must be drawn between steady and variable line 
profiles (e.g., Krolik et al. 1991).  The variable line profile 
$\Delta \mathscr{L}_\nu(t)$ can be defined as 
the component of the total observed line profile $\mathscr{L}_{\nu}(t)$ that 
actually varies in response to continuum fluctuations, while the steady 
line profile $\Lnuavg$ is a time-averaged background component 
(that may or may not correspond to the BLR gas);  symbolically,
\beq \mathscr{L}_{\nu}(t) = \Lnuavg + \Delta \mathscr{L}_\nu(t) .\label{Lnet}\seq
The principle behind reverberation mapping is that the variable line profile, as 
observed at time $t$, is caused by small fluctuations
of the continuum light curve $L_X$ at some earlier time $t - \tau$ 
(typical fractional rms variability amplitudes are $\lesssim 20\%$; e.g., De Rosa
et al. 2015).  Reworded from the standpoint of this paper, 
this principle implies that given the impulse
response function $\Psi(\nu,\tau)$ (i.e. a model of the BLR) and
the light curve of continuum fluctuations, $\Delta L_X = L_X - L_0$ (with
$L_0$ a reference continuum level), we can 
predict the shape of the variable line profile through the convolution
\beq \dL(t) = \int_0^\infty \Psi(\nu,\tau) \dC d\tau. \label{Conv} \seq

Returning to equation \eqref{RHdef}, consider the response 
of the gas to a change in ionizing continuum flux $\Delta F_X$ 
as seen \emph{in the rest frame of the source}, i.e. according to 
an observer located at position $r=0$ in a spherical coordinate system
centered on the BLR.  Then the increased continuum flux,
$\Delta F_X(t'-r/c) = \Delta L_X(t' - r/c)/4\pi r^2$,
received by a gas parcel at time $t'$ and position $r$ is perceived by the observer
to have been emitted by the continuum source at the earlier time $t'-r/c$.
Here we invoked several of the basic assumptions 
used in almost all reverberation mapping studies of the BLR: 
point source continuum emission, 
straight line propagation from source to gas parcel, 
and no plasma effects (ensuring the constant propagation speed $c$).
Provided $\Delta F_X$ is small relative to $\Favg$, the emissivity
can be expanded as 
\beq j_\nu(\Favg + \Delta F_X(t' - r/c)) \approx \javg + \pd{j_\nu}{F_X}\Delta F_X(t' - r/c). 
\label{jexpansion}\seq
By inserting this relationship
into equation \eqref{RHdef} and making a comparison with equation \eqref{Lnet}, 
we identify
\beq \Lnuavg =  \int dV\, \javg \beta_\nu, \seq
and
\beq \dL(t') =  \int dV\, \pd{j_\nu}{F_X}\Delta F_X(t'-r/c) \beta_\nu. 
\label{dLtprime}\seq
The first equation just states that the steady line profile is computed as in equation 
\eqref{RHdef}, 
but in a time averaged sense, while the second equation reveals that 
$\partial{j_\nu}/\partial{F_X}$,
termed the \emph{responsivity}, is fundamental to reverberation mapping.  

Since we are after the luminosity seen by a distant observer,
we need to account for the additional time delay for emitted photons 
to travel from $r$ to the observer plane (i.e. an imaginary plane oriented 
perpendicular to $\hat{n}$ and located beyond the outer 
edge of the emitting volume). 
We must further sum over all times $t'$ that contribute to observed emission 
at the distant observer's time $t$:
\beq \dL(t) = \int dt' \: \dL(t')\: \delta \left[ t - \left(t' - \f{\mathbf{r}\cdot\hat{n}}{c} \right) \right] .
\label{dLt}\seq
Here, all of the basic assumptions listed above were once again invoked,
and we additionally made the (standard) assumption of negligible recombination times
(because these times are typically very short).
Replacing $\Delta F_X$ with $\Delta L_X/4\pi r^2$ in equation 
\eqref{dLtprime} and then substituting equation \eqref{dLtprime} into equation
\eqref{dLt} gives
\beq
\begin{split}
\dL(t) =  & \int dt' \int dV\, \pd{j_\nu}{F_X}\f{\Delta L_X(t' - r/c)}{4\pi r^2} \beta_\nu \\
 & \times \:\delta \left[ t - \left(t' - \f{\mathbf{r}\cdot\hat{n}}{c} \right) \right]  \label{dLt2}.
\end{split}
\seq
The impulse response function is by definition the ratio of $\dL$ to $\Delta L_X$ for a 
delta-function continuum fluctuation,
\beq \Psi \equiv \f{\dL}{\Delta L_X} \: \delta(t'-r/c). \label{TFdef}\seq
Making the substitution ${\Delta L_X} \rightarrow {\Delta L_X}\,\delta(t'-r/c)$ in equation 
\eqref{dLt2} collapses the $dt'$ integral, thereby defining the total time delay
\beq \tau(\mathbf{r}) =  \f{r}{c} \left(1 -  \hat{r}\cdot\hat{n} \right), \label{delay-eqn}\seq
so that the impulse response function can be written as
 \beq
\IRF = \int dV\, \pd{j_\nu}{F_X} \f{\beta_\nu}{4\pi r^2}
 \:\delta [ t - \tau] .
 \label{TFdim}
 \seq
Equation \eqref{TFdim} is seen to be consistent with BM82's equation (2.15).  
Specifically, the responsivity (which has units $\rm{cm}^{-1}~\rm{s}$) is 
analogous to their `reprocessing coefficient' $\varepsilon$, while their factor $g$ 
(the projected 1D velocity distribution function) is unity in the hydrodynamic 
approximation.  The only difference is our inclusion of the escape 
probability $\beta_\nu$ to account for the effects of anisotropy using the 
formalism of RH83.  

\subsection{Responsivity and opacity distributions}
\label{sec:responsivity}
The derivation leading up to equation \eqref{TFdim} is quite general as far as the 
radiative transfer is concerned.  We now specialize to the Sobolev approximation
by following Rybicki \& Hummer (1978) and RH83, in which case
\beq j_\nu(\mathbf{r}) = k\, S_\nu\,\delta\left[\nu - \nu_0 - 
\f{\nu_0}{c}v_l\right], \label{jRH} \seq
where $k = (\pi e^2 / m_e c) f_{12} n_1$ $[\rm cm^{-1}~s^{-1}]$ is the integrated line opacity
of the transition with oscillator strength $f_{12}$ and population number density $n_1$, 
$S_\nu$ is the source function,
$\nu_0$ is the line center frequency, and $v_l \equiv \hat{n}\cdot \mathbf{v}$ is the line of sight velocity
of the emitting gas which has bulk velocity $\mathbf{v}$.  
The delta-function here arises from the use of the Sobolev approximation, 
for when it holds, locally Gaussian line profiles will effectively
behave as delta-functions (see, for example,  \S{8.4} of Lamers \& Cassinelli 1999).  
Note that this statement is not equivalent to our 
assumption that the intrinsic line profile is much narrower than a Gaussian.

The argument of the delta-function accounts
for a non-relativistic Doppler shift only.  There will also be a transverse redshift
that can be of order $1.5(v_t/c)^2\times10^5~\rm{km~s^{-1}}$, where $v_t$ is
the velocity component perpendicular to the LoS, as well as a gravitational redshift
of order $1.5(r_s/r)\times10^5~\rm{km~s^{-1}}$, where $r_s = 2GM_{BH}/c^2$ is the 
Schwarzschild radius.  
Since the PK04 domain extends to a minimum radius $r_{\rm{min}} \approx 30\,r_s$
and the highest velocities in the domain are $\sim 0.1\,c$, either effect
can potentially lead to shifts $\sim\rm{1500~km~s^{-1}}$ at the base of the profile. 
While acknowledging that these are important effects,
we ignore both relativistic redshifts to first order on the grounds that these estimates
are still small compared to the widths of our calculated line profiles
and will apply mainly to the innermost gas, leading to a red wing.

The source function $S_\nu$ in equation \eqref{jRH} describes all radiative processes responsible for
the line emission and in general can be divided into two contributions: 
(i) local intrinsic emission processes, and 
(ii) scattered emission.
We mention below how to realistically model (i), but in this work
we adopt simple scaling relations to account for (i) in a way that will
enable us to compare our results with those from prior works.  
It is known that a proper treatment of (ii) is important when calculating steady line profiles, 
but it is beyond the scope of this work to investigate the importance of scattering for shaping
variable line profiles.

To calculate the variable line profile, we need to 
specify the responsivity, $\partial{j_\nu}/\partial{F_X}$.   
A self-consistent determination of the responsivity requires detailed photoionization
modeling coupled with radiation hydrodynamical simulations.  The former type of 
calculation has been frequently explored without regard to the latter (e.g., 
Dumont \& Collin-Souffrin 1990; Krolik et al. 1991; Goad et al. 1993; 
Korista \& Goad 2000, 2004; Goad \& Korista 2014).  Here we take a first step in 
performing the latter type of calculation.  In \S{\ref{sec:realistic_models}} 
we outline a basic modeling strategy that should be suitable for constraining 
BLR models upon making a comparison with observations.  In essence,
the velocity and density fields are found by performing hydrodynamical 
simulations, and then separately 
the responsivity and opacity distributions are obtained by carrying out
photoionization calculations using the hydrodynamical simulation results as input.
 
For this initial investigation, we opted 
for a simpler approach by adopting prescriptions for the responsivity and opacity
distributions.  To reach a common ground with past investigations, 
we note that it is has been common to adopt a power-law dependence 
for the responsivity (e.g., Goad et al. 1993, 2012)
similar to the one introduced by Krolik et al. (1991), 
who assumed the power can be radially dependent and takes the form
$ \eta(r) \equiv \partial \ln{S_l} / \partial \ln{F_X},$
where $S_l$ is the local brightness of the line-emitting gas.  
Phrased in terms of the source function, this is equivalent to the ansatz
\beq S_\nu(\mathbf{r}) = A F_X^{\eta(r)}, \label{Sfunc}\seq
where $A$ is a function of position, specified below, that sets the overall
response amplitude.  Photoionization modeling 
indicates that $\eta$ typically ranges between 0 and 2 
(see e.g., Krolik et al. 1991; Goad et al. 1993, 2012).
For simplicity, we adopt $\eta = 1$ in this work, which gives $A$ units
of seconds and defines our responsivity as 
\beq \pd{j_\nu(\mathbf{r})}{F_X} = k\,A\,\delta\left[\nu - \nu_0 - \f{\nu_0}{c}v_l\right].
 \label{responsivity} \seq

Specifying the magnitude of $A$ is only necessary when making
quantitative comparisons with observed spectra.  We will use
arbitrary flux units, allowing the constant $A_0$ in our fiducial
relation, 
\beq A(r) = A_0 (r/r_1)^2,\seq
where $r_1$ is one light day, to serve as a normalization factor.
Our results are calculated using this heuristic prescription for $A(r)$, 
which we motivate below, although in \S{\ref{sec:disk_only}} we present 
an example calculation with $A(r)  = A_0$ instead.    

To obtain an expression for the responsivity that involves only hydrodynamical quantities, 
we estimate the number density of the lower level of the transition in question
in terms of the fluid density $\rho$ through
\beq n_1(\mathbf{r}) = \Ae\xi_{\rm ion} \f{\rho}{\mu m_p}, \seq
where $\xi_{\rm ion}$ is the ion fraction of the emitting ion with elemental abundance
$\Ae$, and $\mu$ and $m_p$ are the mean molecular weight and mass of a proton, respectively.  
 These quantities are assumed to characterize the state of the gas \emph{after} the change 
in photoionizing flux.  We can now define an effective opacity per unit mass as
\beq \kappa =  \left( \f{\pi e^2}{m_e c} \right) \f{\Ae \xi_{\rm ion} f_{12}}{\mu m_p \nu_0} ~~[{\rm cm^2\,g^{-1}}],\seq
and in our calculations we take $\kappa$ to be a spatially fixed quantity throughout the domain.
Note that in writing equation \eqref{responsivity} we have assumed that the flux dependence
of the emissivity is dominated by that of the source function,
i.e. that $k = \kappa \rho \nu_0$ is insensitive to changes in the ionizing flux.  
This will not be true in general since $\kappa$ depends on the ion fraction, while hydrodynamic effects can lead to changes in $\rho$.
Ignoring the latter possibility (since it implies a nonlinear response; see \S{\ref{sec:realistic_models}})
therefore implies that $k$ is independent of $F_X$ when $\kappa$ is treated as a constant.

As a very simple example of what the above scalings imply, consider a spherically symmetric, 
\emph{constant}, high-velocity outflow illuminated by an isotropic source at its center.  By mass
conservation, the density scales as $r^{-2}$, and therefore so does $k$.  Then
 $A \propto r^2$ amounts to assuming that the emissivity of the gas is directly 
 proportional to the density, while the responsivity 
 ($\partial j_\nu/\partial F_X \propto \kappa \rho r^{2}$) 
is constant with radius 
since the emissivity and flux both falloff as $r^{-2}$.  In contrast, taking $A = A_0$
implies $j_\nu \propto r^{-4}$ and $\partial j_\nu/\partial F_X \propto r^{-2}$; 
this scaling reproduces the results of Chiang \& Murray (1996), 
as shown in the Appendix.

\subsection{The escape probability}
\label{sec:esc_prob}
In equation \eqref{TFdim}, the escape probability, assuming a single resonant surface,
is given by (RH83)

\beq \beta_\nu(\mathbf{r}) = \f{1-e^{-\tau_\nu} }{\tau_\nu}. \label{beta}\seq
Treating multiple resonant surfaces, which can arise for non-monotonic velocity fields,
modifies equation \eqref{beta} by an additional multiplicative factor of $e^{-\tau_\nu}$
for each surface,
but we expect equation \eqref{beta} to capture the dominant optical depth effects.
In the Sobolev approximation, the optical depth is given by
\beq  \tau_\nu(\mathbf{r}) = \f{k}{\nu_0} \f{c}{|dv_l/dl|}, \seq
where $dv_l/dl \equiv \hat{n} \cdot \nabla v_l$ is the 
line of sight velocity gradient, often denoted as $Q$: 
\beq
\f{dv_l}{dl} \cong Q(\mathbf{r}) =\sum\limits_{i,j} \frac{1}{2} \left( \frac{\partial v_i}{\partial r_j} +\frac{\partial v_j}{\partial r_i} \right).
\label{eq:Q_sobolev}
\seq
The components of $Q$ in various coordinate systems can be found in Batchelor (1967).  
Therefore, the product $k \,\beta_\nu$ present in the integrand of 
equation \eqref{TFdim} can be written
\beq k(\mathbf{r}) \,\beta_\nu(\mathbf{r}) =  \f{\nu_0}{c}\left|\f{dv_l}{dl}\right| (1-e^{-\tau_\nu}). \label{kappabeta}\seq
Notice that this product is only dependent on the density and opacity
through the optical depth.  For $\tau_\nu \gg 1$, this dependence is very weak and the escape of 
photons is primarily governed by the local LoS velocity gradient.  
Once $\tau_\nu \lesssim 0.1$, on the other hand, $\beta_\nu \approx 1 - \tau_\nu/2$, and 
 the impulse response function becomes weakly dependent on $|dv_l/dl|$, instead  
depending primarily on the magnitude of $k$ (i.e. the product of the density and opacity),
which must be smaller than $(\nu_0/c) |dv_l/dl|$.  Thus, in general, the response will be weaker
for reprocessed photons emitted in an optically thin region compared to an optically thick, 
rapidly moving region. 

\subsection{The resonance condition}
\label{sec:res_cdn}
Having derived formulae for the quantities appearing in the
integrand of equation \eqref{TFdim}, we can express the impulse response 
function in spherical coordinates as 
\beq
\begin{split} 
\Psi(y,t) &= \f{1}{4\pi c}  \int_{r_{\rm{in}}}^{r_{\rm{out}}}dr \int_0^\pi \sin\theta d\theta \int_0^{2\pi} d\phi\, A(r) \\
&\times \left|\f{dv_l}{dl}\right| (1-e^{-\tau_\nu}) 
\: \delta\left[y- v_l' \right]
\: \delta \left[ t - \tau \right],
 \label{IRF}
\end{split}
\seq
where $r_{\rm{in}}$ and $r_{\rm{out}}$ are the inner and outer radii of the reverberating region
and we have defined the dimensionless frequency shift 
$ y \equiv (\nu - \nu_0)/\nu_0$ and denoted $v_l/c = v_l'$.
The argument of the first delta function defines an iso-frequency surface
specifying all physical locations that contribute to a given frequency  shift $y$.
Likewise, the argument of the second delta function defines an iso-delay surface, 
giving all points in the volume with nonzero responses at a given time $t$.  
Only the intersection of these two surfaces contribute to the integral at a given $(y,t)$.
We will refer to locations satisfying the combined arguments as resonance points, 
and to the equation governing these locations as the resonance condition.  
 
For axisymmetric models, to which we confine ourselves to in this work, the resonance
condition is used to solve for the resonant azimuthal angles $\tilde{\phi}$ corresponding
to each $(r,\theta)$ coordinate on the grid. 
It is clear that dependence on $\phi$ enters through $\hat{n}$.  Two angles are
required to specify $\hat{n}$, namely the observer's azimuthal and polar coordinates $(\phi_n,\theta_n)$.
Without loss of generality we choose $\phi_n =0$, while $\theta_n$ is the same as the LoS inclination angle,
hereafter denoted $i$.  Then the components of $\hat{n}$ are 
$n_r =  \sin\theta \cos\phi \sin i + \cos\theta \cos i$,
$n_\theta =  \cos\theta \cos\phi \sin i - \sin\theta \cos i $, and
$n_\phi = - \sin\phi \sin i $, 
giving for the resonance condition the coupled algebraic equations
\beq
\begin{split}
y = & \, n_r v_r'(r,\theta) + n_\theta v_{\theta}'(r,\theta)  + n_\phi v_{\phi}'(r,\theta); \\
t = & \, (r/c) \left(1 - n_r \right).
\label{resonance_cdn}
\end{split}
\seq
Here the primes on the velocity components indicate that they are in units of $c$
(consistent with our convention for $v_l'$ above).
For \emph{analytic} axisymmetric hydrodynamic solutions, equations 
\eqref{resonance_cdn} can be easily solved for $\phi = \tilde{\phi}$, given $y$, $t$, 
and $i$.  However, there is a subtlety that arises for discretized solutions, requiring
first the solution of an alternate form of the resonance condition, 
equation \eqref{sketch} below.  We return to this point and discuss our actual
procedure in \S{\ref{sec:methods_numerical}}.

\begin{figure*}                            
\centering
\includegraphics[width=0.54\textwidth, trim=0 0 0 0, clip=true]{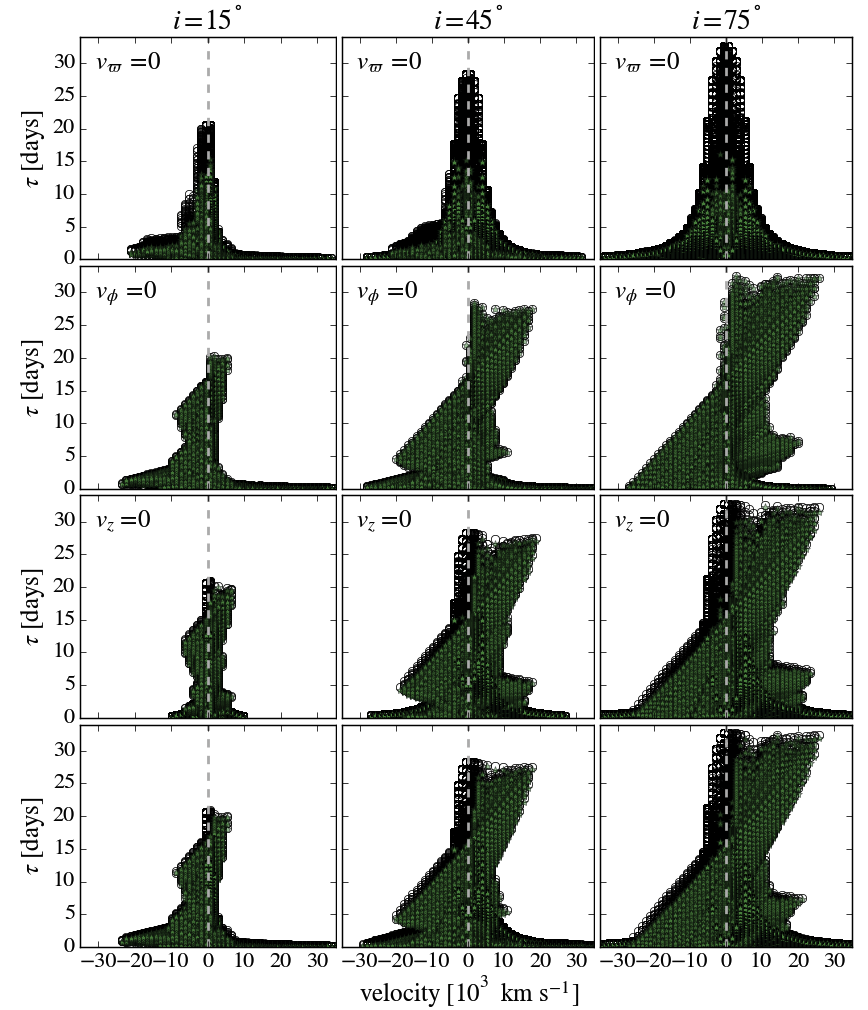}
\raisebox{0.9in}{\includegraphics[width=0.235\textwidth, trim=0 0 0 0, clip=true]{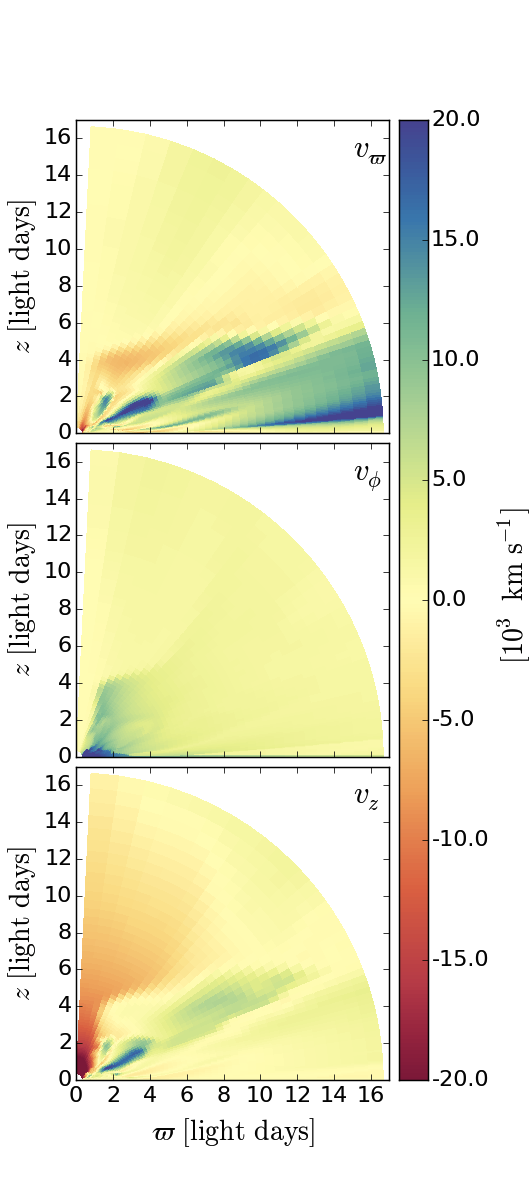}} 
\caption{Echo image sketches of the PK04 solution for $i = 15^\circ$ (1st column), $i = 45^\circ$ (2nd column), 
and $i = 75^\circ$ (3rd column).  These are plots of the two time delays, 
$t_+$ and $t_-$ (green and black symbols, respectively, but note $t_+ = t_- $ when $v_\phi = 0$), 
corresponding to each LoS velocity, 
found by solving equation \eqref{sketch} using the velocity components from the 
PK04 solution.  The last column displays maps of these velocity components.  
The first three rows of echo image sketches
shows the effect of zeroing the (cylindrical) velocity component shown in the 
corresponding map.  For example, the 1st row of sketches has nonzero $v_\phi$ and $v_z$.
The sketch for $i = 75^\circ$ in this row shows a characteristic `virial envelope', which is due to $v_\phi$ alone;
at lower inclinations contributions from $v_z$ become visible.  
In the 2nd row of sketches there is no virial envelope, as only the poloidal velocity components are nonzero;
comparison with the 1st row reveals that the diagonal features are caused by $v_\varpi$.  
Vertical dashed lines are plotted at line center to highlight an overall blue-shift effect that is absent 
in the 3rd row, which has $v_z = 0$ and hence lacks any shift caused by $v_z \cos i$ 
in equation \eqref{vzcosi}.  This effect is best seen by comparing the bottom
row of sketches, which accounts for the full PK04 velocity field, with the 3rd row.  We emphasize that
these sketches can be used to assess where an echo image \emph{cannot} show a response, 
but elsewhere they need not resemble the actual image since $\Psi(y,t)$ may be negligible.  
 }
\label{fig:sketches}
\end{figure*}
  
\subsection{Echo image sketches}
\label{sec:sketches}
Welsh \& Horne (1991) derived simple equations relating the 
velocity field and the time delay for specific outflow, inflow, and 
Keplerian velocity fields, which allowed them to sketch 
velocity vs. delay and thereby show the \emph{possible outlines} of 
echo images.  A general equation for `echo image sketches'
of axisymmetric models is found by   
eliminating $\phi$ from equations \eqref{resonance_cdn};
it is simplest to write down using cylindrical velocity components, 
($v_\varpi, v_\phi, v_z$):
\beq
\begin{split}
t = \f{r}{c} \Biggr[1 -
& \cos\theta\cos i - \f{\sin\theta}{v_\varpi'^2 + v_\phi'^2} \\ \times\Biggl( v_\varpi' y'  
& \pm v_\phi' \sqrt{(v_\varpi'^2 + v_\phi'^2)\sin^2i - y'^2} \Biggr)
\Biggl]
\label{sketch},
\end{split}
\seq

where 
\beq y' \equiv y - v_z' \cos i .\label{vzcosi}\seq 
Equation \eqref{sketch} reduces to the simpler ones presented in 
Welsh \& Horne (1991), i.e.   
the relationship for a spherical inflow/outflow is
obtained by setting $\theta = -\pi/2$
and $v_\phi = v_z = 0$, giving 
\beq t = \f{r}{c}\left[ 1 + \f{y}{v_\varpi'} \right], \seq
whereas a Keplerian disk satisfies,
\beq \left[\f{t - r/c}{r/c}\right]^2 + \left[ \f{y}{v_\phi'} \right]^2 = \sin^2 i, \seq
obtained by setting $\theta = \pi/2$ and $v_\varpi = v_z = 0$.  

Figure 1 shows echo image
sketches for the PK04 solution.  From top to bottom, the first three rows show the effect
of zeroing each velocity component, maps of which are plotted in 
the right column.  The top row lacks the prominent diagonal feature
present in the other rows, indicating that it is due to the $v_\varpi$ component.  
Note that diagonal features are expected for radial outflows (c.f. Welsh \& Horne 1991).

The final row shows sketches with all velocity components nonzero.  
A comparison with the third row highlights a tendency for echo images
of outflows to exhibit blue-shifted excesses.  This effect is clearly revealed by
equation \eqref{vzcosi}: the velocity shift $y = (\nu - \nu_0)/\nu_0$ is
offset by a factor of $(v_z/c) \cos i$, so the vertical velocity component
causes a blueshift for positive $v_z$ and a redshift for negative $v_z$.
This will only be significant at small inclinations ($i \lesssim 45^\circ$) due 
to the factor of $\cos i$.  We will examine this result more closely in 
\S{\ref{sec:i_dependence}}.

The significant differences between the bottom and top rows of sketches 
hints that it may be possible to infer the presence of
a poloidal velocity field through observations of echo images.  
However, these sketches are mainly useful for visualizing the mapping
from physical space to velocity-delay space, thereby showing 
which regions of an echo image \emph{cannot} show a response.  
Most of the features outside of the `virial envelope' formed by the rotational
velocity component turn out to have much smaller fluxes unless the lines
originating in the wind are very optically thick.

\subsection{Transfer~functions and line~profiles} 
\label{sec:definitions}
Most reverberation mapping studies to date have primarily focused on two 
quantities derived from the impulse response function.  
The first is the transfer function, 
which is the frequency-integrated impulse response function,
\beq
\Psi(t) = \int_{-\infty}^\infty \Psi(y,t)  \, dy . \label{IRF-TF}
\seq
In practice, the transfer function is the quantity used to calculate 
mean time lags, and hence to measure a characteristic radius of the BLR.  
Similarly, we can also define the line profile by
\beq
\Phi(y) = \int_0^\infty \Psi(y,t) \, dt , \label{IRF-LP}
\seq
where the limits are $(0,\infty)$ since $\Psi(y, t  < 0) = 0$.  
Note that $\Phi(y)$ is not the same as the variable line profile defined in
equation \eqref{Conv}.  Rather, it is (to within a normalization factor) the 
limiting case of a variable line profile found by convolving $\Psi(y,t) $ 
with a \emph{constant} continuum light curve.  As such, the 
line profiles presented in this paper should be viewed as merely
representative of the line shapes expected for our disk wind models.   
Detailed predictions of variable line profiles are system
specific, as they require carrying out the convolution with the
observed continuum light curve $\Delta L_X(t)$. 

\section{Methods}
\label{sec:methods}
Two approaches for calculating impulse response functions
from models of the BLR were introduced early on.  A stochastic
approach was taken by Welsh \& Horne (1991) and 
P\'erez et al. (1992), in which a domain was populated
with a large number ($\sim760,000$ and $25,000$, respectively) 
of points, satisfying some assigned velocity field, spatial distribution,
and emissivity.  These discrete particle models continue to provide 
intuition into the nature of the mapping between physical space and 
frequency-delay space.

An analytic approach was taken by BM82 and later by
Chiang \& Murray (1996; hereafter CM96), whose
BLR model consisted of an axisymmetric Keplerian disk combined 
with a simple radial wind prescription.  Here we adopt CM96's approach,
extending it to allow the exploration of both 2-D analytic and 
numerical hydrodynamical models.

\subsection{Formal evaluation of the impulse response~function}
Simplifying equation \eqref{IRF} to its basic functional form and 
changing integration variables to $\mu \equiv \cos\theta$ gives
\beq
\Psi(y,t) =  \int_{r_{\rm{in}}}^{r_{\rm{out}}}dr \int_{-1}^1 d\mu \int_0^{2\pi} d\phi\, I 
\: \delta\left[y - v_l'\right]
\: \delta \left[ t - \tau \right],
 \label{IRFbasic}
\seq
where 
\begin{equation*}
I(\mathbf{r}) = \f{1}{4\pi c} A(r) \left|\f{dv_l}{dl}\right| (1 - e^{-\tau_\nu}).
\end{equation*}
To make use of the delta functions, any pair among 
$(dr,d\mu)$, $(dr,d\phi)$, and $(d\mu,d\phi)$ 
can be replaced by $(dv_l',d\tau)$ using a Jacobian.
For axisymmetric problems in which the density and velocity fields are independent of $\phi$, 
it is natural to replace either $(dr,d\phi)$ or $(d\mu,d\phi)$, so that the triple 
integral can be reduced to a single integral over $\mu$ or $r$.  
To make a clear comparison with CM96, 
we chose to use $(dr,d\phi)$, so the mapping reads
\beq dr \, d\phi \, |J| = dv_l' \, d\tau, \label{Jacobian} \seq
where
\beq |J(\mathbf{r}) | = \left| \left(\pd{\tau}{r} \right) \left(\pd{v_l'}{\phi} \right) - \left(\pd{\tau}{\phi} \right) \left(\pd{v_l'}{r} \right) \right| .\seq
Equation \eqref{IRFbasic} becomes
\beq
\Psi(y,t) =  \int_{-1}^1 d\mu \int dv_l' \int d\tau \, \f{I}{|J|} 
\: \delta\left[y - v_l'\right]
\: \delta \left[ t - \tau \right],
 \label{IRFdeltas}
\seq
which evaluates to
\beq
\Psi(y,t) =  \int_{-1}^1 d\mu \, \left[ \f{I}{|J|} \right]_{(\tilde{r}, \mu, \tilde{\phi})}.
 \label{IRFformal}
\seq
We use the subscript notation to indicate that for each $\mu$, the 
integrand is to be evaluated at the resonance point 
$(\tilde{r},\tilde{\phi})$ corresponding to a given $(y,t)$; geometrically this point
will lie somewhere in a conical slice $(r,\phi)$ through the volume. 
Its location is determined by the solution to 
the resonance condition, equation \eqref{resonance_cdn}.
Assuming motion purely in the midplane ($\mu = 0$), CM96's result 
can be obtained with the substitution
$I \rightarrow I\,\delta[\mu - 0]$, as we illustrate in the Appendix.

\subsection{Numerical evaluation of the impulse response~function}
\label{sec:methods_numerical}
To numerically evaluate the remaining integral over $\mu$,
we employ the trapezoid rule, leading to the discrete form
\beq \Psi(y,t) \approx \f{1}{2}\sum_{k=1}^{N-1} \Delta\mu_k 
\left[ \left.\f{d\Psi}{d\mu}\right|_{k+1} + \left.\f{d\Psi}{d\mu}\right|_k \right], \label{trap_rule}\seq
where we have used the simplifying notation
\beq \f{d\Psi}{d\mu} = \left[ \f{I}{|J|} \right]_{(\tilde{r}, \mu, \tilde{\phi})}. \label{integrand}\seq 
Note that for grid-based simulation data in spherical coordinates, 
the native grid spacing can be used to arrive directly at
$\Delta \mu_k = \mu_{k+1} - \mu_k$.  (Otherwise, the discretized solution 
would need to be interpolated to a spherical grid or a different Jacobian 
would need to be defined.)

As mentioned in \S{\ref{sec:res_cdn}}, when applied to simulation data, 
a subtlety arises in the evaluation of the integrand,
equation \eqref{integrand}.
To clarify what is involved, it should first be emphasized that 
the goal is to arrive at a legitimate digital image to compare
with echo images obtained from observations.  That is, we 
need to construct a 2-D array of pixels with the center of 
each pixel at specified values of $(y,t)$, 
and the magnitude of $\Psi(y,t)$ determining
the value of the entire pixel.
Ideally, we would like to directly evaluate each of the
$N$ values of $d\Psi/d\mu$ precisely at $(y,t)$.  
However, this cannot be done in practice.  
The reason is that with discretized data, it is 
impossible to find resonance points exactly at the
center locations of pixels to an acceptable tolerance level.  
Indeed, as equation \eqref{sketch} reveals, there are only
certain values of $y$ that satisfy the resonance condition for
a given $t$, and vice versa, \emph{when the grid coordinates 
($r,\mu$) and velocity fields are given}.

Our procedure to generate an echo image therefore involves
interpolating from the resonant locations nearest the center 
of each pixel.    
For every value of $y$, i.e. for every column of pixels in our 
image array, we loop through all grid points of our simulation 
and associate each one with a specific value of $t$ that satisfies 
equation \eqref{sketch}.  We do the same for each row of
pixels, collecting all $y$ values that correspond to a given $t$.
For each pixel, we then evaluate 
$d\Psi(y_L,t)/d\mu$, $d\Psi(y_R,t)/d\mu$,
$d\Psi(y,t_A)/d\mu$, and $d\Psi(y,t_B)/d\mu$, 
where $(y_L,t)$, $(y_R,t)$, $(y,t_A)$, and $(y,t_B)$ are the four 
locations nearest to (i.e. left of, right of, above, and below, respectively) 
the center of the pixel.   
Lastly, we bilinearly interpolate the four values of $d\Psi/d\mu$
to arrive at $d\Psi(y,t)/d\mu$.  By adding up all such values of 
$d\Psi(y,t)/d\mu$ in accordance with equation \eqref{trap_rule},
we finally arrive at $\Psi(y,t)$, whose magnitude is assigned
to that pixel.  

\subsection{Direct vs. indirect calculation of the transfer~function and line profile}
\label{sec:LPTFcalc}
If provided with an analytic hydrodynamical model (e.g., that of CM96), there
is no need to carry out the interpolation procedure just described,
since resonance points can be found for any $(y,t)$.
By summing over the rows and columns of resulting echo image 
with a suitable algorithm such as the trapezoid rule, 
excellent numerical approximations to the integrals in equations 
\eqref{IRF-TF} and \eqref{IRF-LP} can be obtained.  We refer to
this method of calculating the transfer function and 
line profile as an indirect one, since it first involves calculating $\Psi(y,t)$.

This summation can also be carried out for discretized solutions, 
using the non-interpolated values of $d\Psi/d\mu$.  However, again
a subtlety arises, which is not easily dealt with.  The issue is the 
double-valued nature of the mapping from $(r,\mu)$ to $(y,t)$.  
From equation \eqref{sketch}, we see that in general there can be two values 
of $t$ for every $y$.  Each will have a different resonant $\tilde{\phi}$ coordinate, as 
they physically correspond to emission regions on opposite sides of the BLR
that have the same time delay.  However, they manifest as separate branches
in a plot of $\Psi(y,t)$ vs. $t$, and we find that one branch (corresponding
to gas on the far side of the BLR) is sampled much less 
densely than the other (due to the logarithmic grid spacing).  
Hence, special integration routines are necessary to accurately
carry out this indirect method, which will be needed to calculate convolutions with
observed light curves; they will be presented in a separate 
paper focused on making a comparison with observations.
 
The direct method for calculating line profiles and transfer functions
is to carry out the integrals over $y$ and $t$ in equations \eqref{IRF-TF} and 
\eqref{IRF-LP} analytically.  Using the impulse response function
in the form of equation \eqref{IRFdeltas}, we find, after some manipulation
of the Jacobian defined in equation \eqref{Jacobian}, 
\beq
\begin{split}
 \Psi(t) & =  \int_{r_{\rm{in}}}^{r_{\rm{out}}}dr \int_{-1}^1 d\mu \: 
 \sum_{i=1}^2  \left[ \f{I}{\left| d\tau/d\phi \right|} \right]_{(r,\mu,{\phi_{t_i}})};\\
 \Phi(y)  & =  \int_{r_{\rm{in}}}^{r_{\rm{out}}}dr \int_{-1}^1 d\mu  \: 
  \sum_{i=1}^2 \left[ \f{I}{\left| dv_l/d\phi \right|}\right]_{(r,\mu,{\phi_{y_i}})}.
 \label{LPandTF}
 \end{split}
\seq
The subscript notation here indicates that the integrands 
are to be evaluated at the location where $t = \tau(r,\mu,\phi)$ in the case
of $\Psi(t)$ and $y = v_l(r,\mu,\phi)$ in the case of $\Phi(y)$; in general
there can be two such locations, $\phi_{t_1}$ and $\phi_{t_2}$ for $\Psi(t)$,
and $\phi_{y_1}$ and $\phi_{y_2}$ for $\Phi(y)$, hence the summations.
We numerically evaluate these integrals (again using the trapezoid rule).
For the technical reasons described in the previous paragraph, 
our results only employ this direct method.  
Nevertheless, we draw attention to the fact that this and the indirect method
are completely independent and therefore provided a useful means to 
benchmark the code used in this work (see the Appendix).  
   
 \section{Results}
\label{sec:results}
The above methods were implemented as a post-processing routine and 
applied to the line-driven disk wind solution presented in PK04.  
The PK04 solution is a hydrodynamic model of an outflow launched
from a geometrically thin,
optically thick disk accreting 
onto $10^8~M_{\odot}$ non-rotating SMBH at a rate of 1.8\msyr.  
For an accretion efficiency $\eta = 0.06$, 
this corresponds to a disk luminosity 
$L_D = 0.5\,$\Ledd, where \Ledd~is the Eddington luminosity.
The numerical setup is similar to that developed by Proga et al. (2000):
for simplicity and to reduce the computational time it
was assumed that X-rays and all ionizing photons are emitted by the
central object, which in term was approximated as a point source.
Specifically, the central engine has $L_X = 0.05 \,$\Ledd~and does not contribute
to the radiation force acting on the wind.  

An important feature of the PK04 solution that indirectly 
contributes to the line-driving mechanism is self-shielding by
the disk atmosphere: 
dense clumps (a ``failed wind'') form at small radii 
as a result of over-ionization, which shield the gas 
launched at large radii from ionizing radiation. 
The resulting line-driven disk wind 
is very fast ($\sim 10^4 ~\rm{km}~{\rm s}^{-1}$) at low latitudes,
where it is directed primarily in the radial direction at small heights
above the disk (see the top panel in Figure 1).  
At somewhat greater heights, namely for $55^\circ \lesssim \theta \lesssim 70^\circ$,
the vertical component of the wind velocity also becomes large. 
It has been shown that this model can produce features observed 
in X-ray spectra of AGN (Schurch et al. 2009; Sim et al. 2010).

The PK04 simulation was performed on a logarithmic grid with resolution 
$[N_r,N_\theta] = [100, 140]$.  
 Our results are calculated assuming emission from only the top half of the disk;
 the bottom half is assumed to be blocked by the disk.  Additionally, 
 we exclude the polar region from $\theta = 0^\circ$ to $8^\circ$, which is very hot
 and optically thin.  Thus, it will typically show negligible response as it hosts few lines.       
 Recall from \S{\ref{sec:responsivity}} that the optical depth can be parametrized in terms of
 the opacity per unit mass $\kappa = k/(\rho \nu_0)$, giving
 $\tau_\nu = c \rho \kappa / |dv_l/dl|$.  We will explore the dependence on $\kappa$
 in \S{\ref{sec:kappa_dependence}}, but elsewhere we adopt a fiducial value of 
 $\kappa = 10^4 \kappa_{\rm{es}}$ with
 $\kappa_{\rm{es}} = 0.4 \: [\rm cm^2 g^{-1}]$ the electron scattering opacity.
 
\begin{figure}                            
 \centering
\includegraphics[width=0.45\textwidth, trim=0 0.5cm 0 0, clip=true]{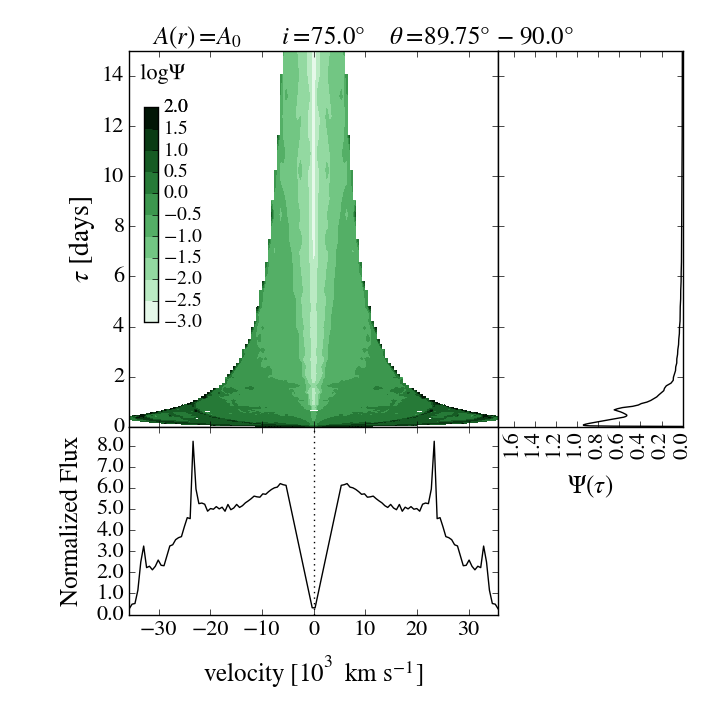}
\includegraphics[width=0.45\textwidth, trim=0 0 0 0.5cm, clip=true]{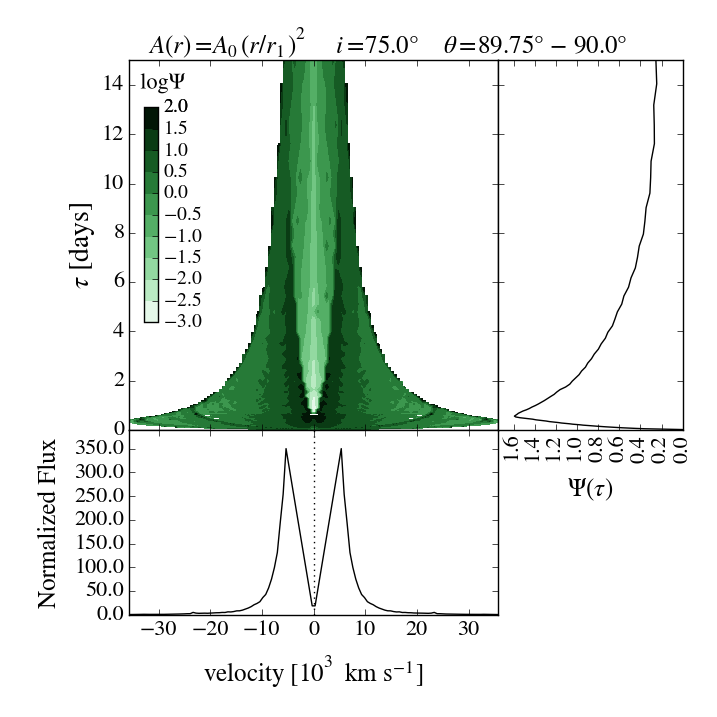}
\caption{Two `disk only' calculations for $i = 75^\circ$ performed by integrating 
over the PK04 solution only in the range $\theta = 89.75^\circ - 90^\circ$.
Top: $A(r) = A_0$. Bottom: $A(r) = A_0 (r/r_1)^2$.  The normalization factor 
$A_0$ is chosen to satisfy $\int \Psi(\tau)\: d\tau = 1$ for the transfer function
in the top panel.
Both cases result in double peaked line profiles, consistent with expectations,
with the lack of line center flux clearly visible in the echo images. 
The prominent spikes  on the line profile at $\pm 23\times10^3~\rm{km\,s^{-1}}$ 
in the top plot coincide with a dark ring on the echo image.  They
are also present in the bottom plot but are masked by the emission from outer
radii caused by the extra $r^2$ dependence in the responsivity.  
 }
\label{fig:disk_case}
\end{figure}
 
\begin{figure}                            
  \centering
\includegraphics[width=0.45\textwidth, trim=0 0.5cm 0 0, clip=true]{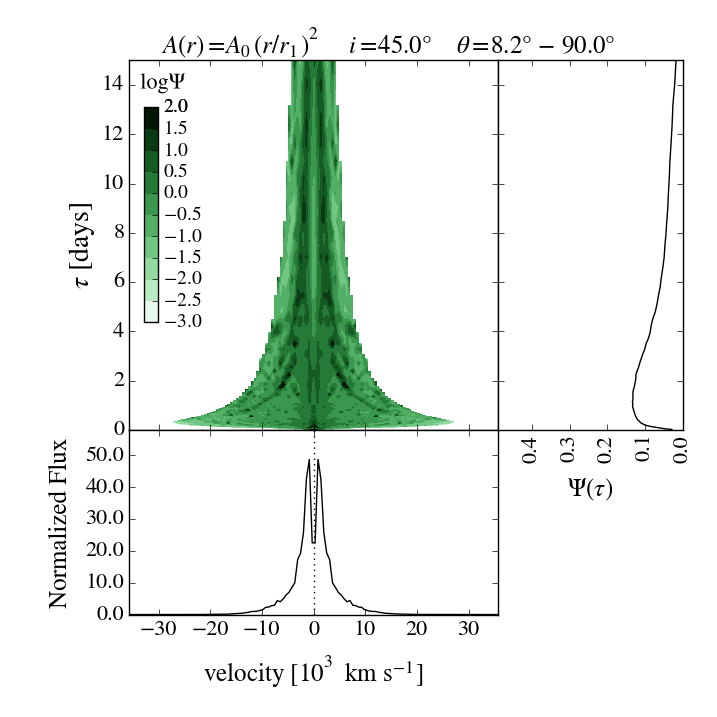}
\includegraphics[width=0.45\textwidth, trim=0 0 0 0.5cm, clip=true]{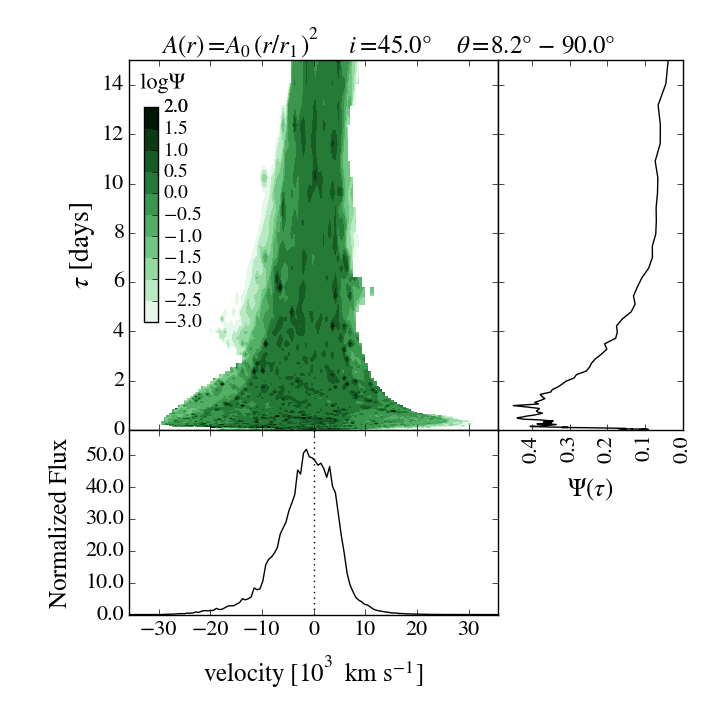} 
\caption{Two disk wind calculations for $i = 45^\circ$ performed by integrating over 
the PK04 solution in the range $\theta = 8.2^\circ - 90^\circ$.
Top: purely rotational case calculated by zeroing all quantities involving $v_r$ and $v_\theta$.
Bottom: full velocity field case.  The normalization factor $A_0$ is again chosen to satisfy
$\int \Psi(\tau)\: d\tau = 1$ for the top transfer function.
The purely rotational case resembles that of the disk-only calculation with $A(r) = A_0 (r/r_1)^2$: 
symmetric echo image, double-peaked line profiles, and
a transfer function displaying an extended-response.  Including the poloidal velocity field 
(i) changes the line profile from double to single peaked; (ii) broadens the line profile
overall; and (iii) results in a blue-shifted excess.   
 }
\label{fig:diskwind_case}
\end{figure}
  
\begin{figure*}                            
\centering
\includegraphics[width=0.43\textwidth, trim=0 0.15cm 4.5cm 0, clip=true]{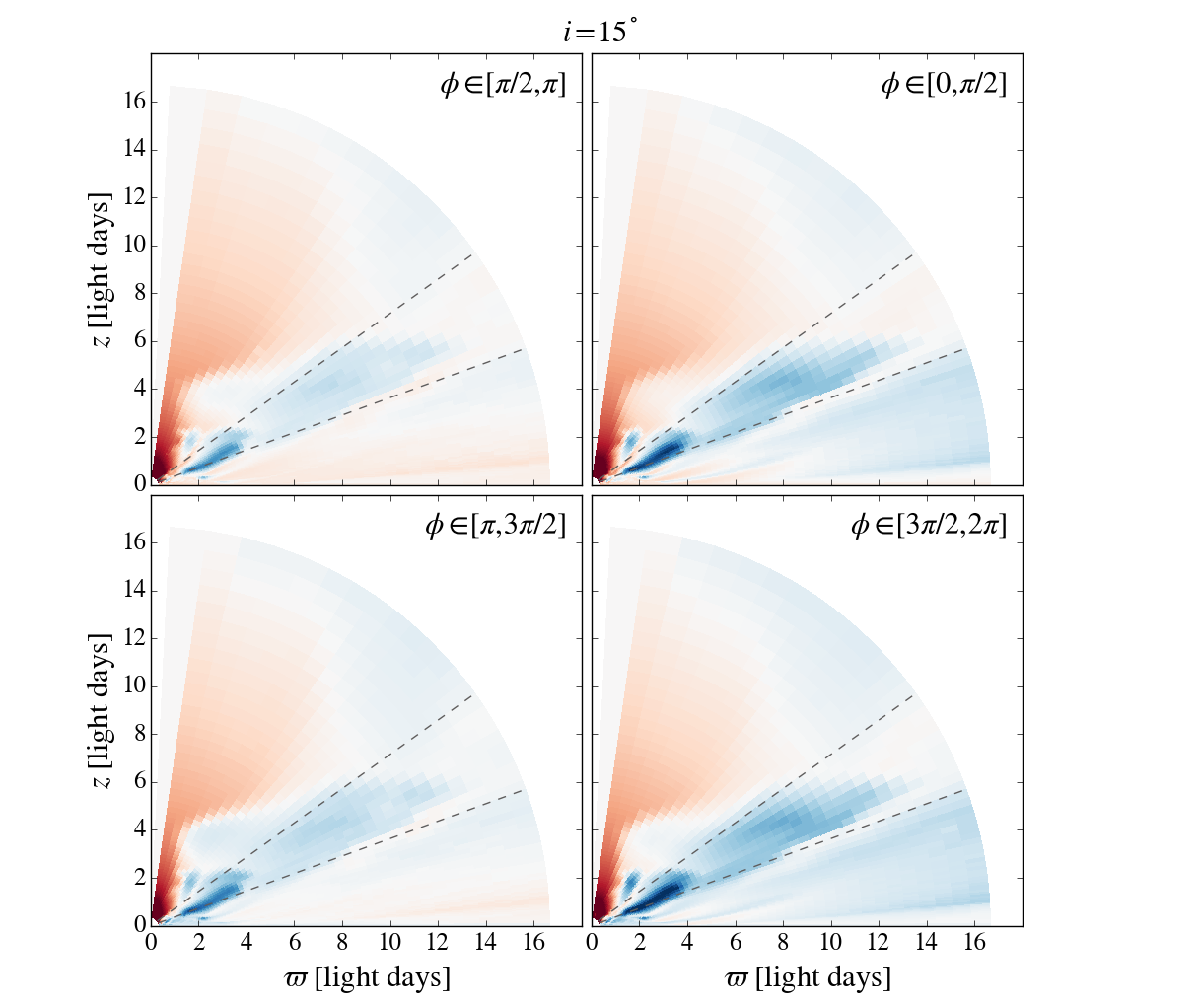} 
\includegraphics[width=0.48\textwidth, trim=3.75cm 0 0 0, clip=true]{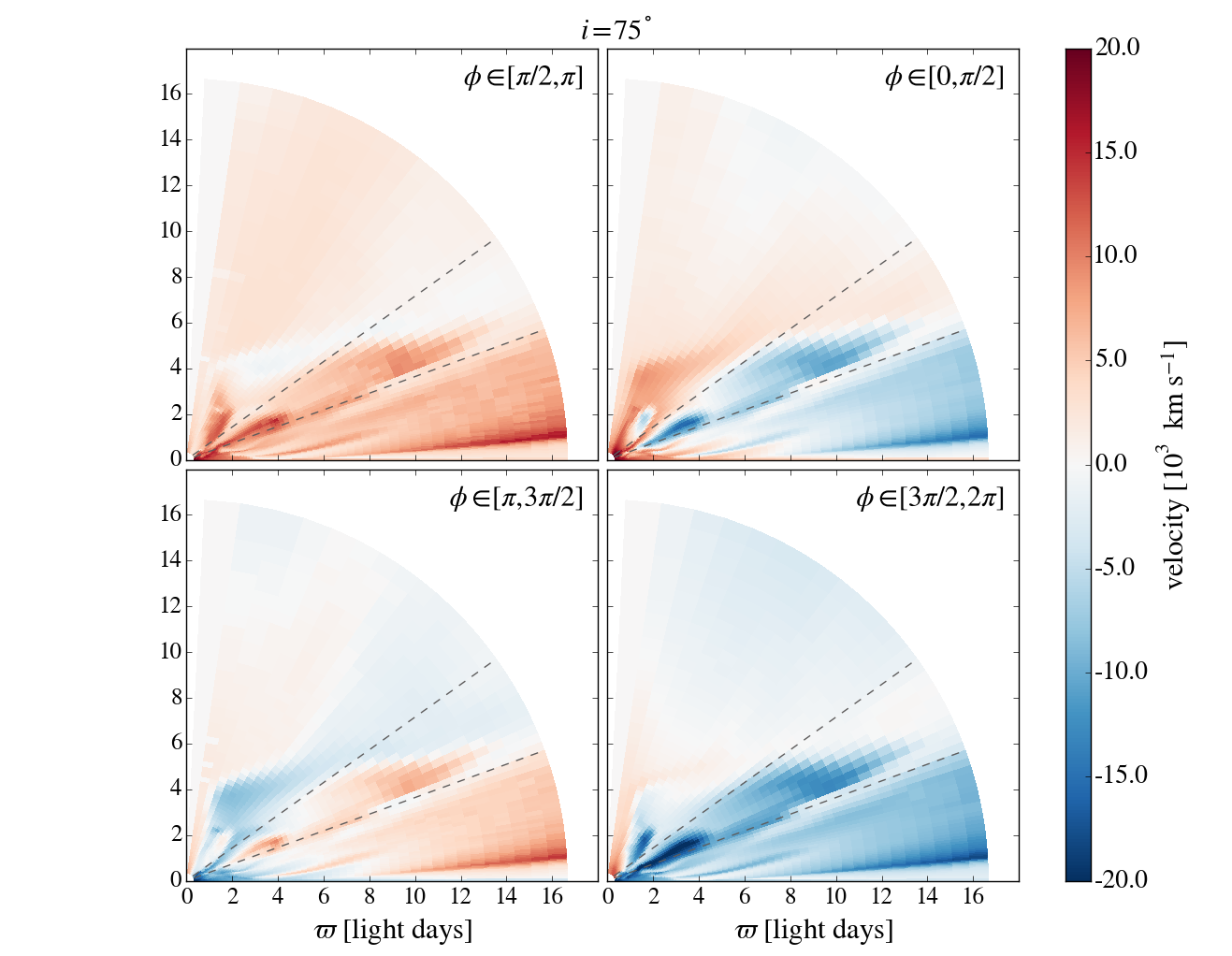}
\caption{Maps of the velocity seen by the observer (i.e.\emph{ negative} $v_l$), averaged over each quadrant of $\phi$, in the PK04 solution for $i = 15^\circ$ (left) and $i = 75^\circ$ (right).   The observer is located at $\phi = 0$, so in each case the right two maps 
represent the front side of the BLR and the left two the backside.  Shades of blue denote regions with $v_l > 0$,
indicating that the gas is blue shifted and moving toward the observer, while red shades denote receding gas that will
contribute to the red side of line profiles.  The portion of the domain with
$55^\circ \lesssim \theta \lesssim 70^\circ$, delineated by the dashed lines, is the region of the PK04 solution with a substantial 
positive $v_z$ component.  It thus always appears blue to the $i = 15^\circ$ observer, but on the far side of 
the BLR it appears red to the $i = 75^\circ$ observer.  This implies that echo images and line profiles can acquire 
noticeable blue-shifted excesses at small inclinations, as Figures \ref{fig:LPTF_istudy} - \ref{fig:kappa_study} reveal.  
Refer to \S{\ref{sec:i_dependence}} for details.
 }
\label{fig:vlos_maps}
\end{figure*}
 
 \subsection{Keplerian disk (no wind): \\
Effects of varying the responsivity}
\label{sec:disk_only}
We begin by analyzing a familiar case: an optically thick, Keplerian disk,
which is expected to show double-peaked line profiles due to a lack of flux 
at line center.  This calculation was done by only integrating from $\theta = 89.75^\circ$
to $\theta = 90^\circ$, an interval that comprises about one third of the 140 grid indices
due to the logarithmic PK04 grid, so there is ample resolution.
While this region constitutes the base of the wind, the poloidal
velocity components are very small relative to the azimuthal component; 
to strictly focus on the Keplerian velocity field, we 
set $v_r$ and $v_\theta$ as well as their gradients to zero.   
Our objective here is to compare the differences between a responsivity 
that is only proportional to the density, implying $A(\mathbf{r}) = A_0$,
and one following our fiducial scaling with $A(r) = A_0 (r/r_1)^2$.     
The results are shown in Figure~\ref{fig:disk_case}.  These plots are 
akin to those presented by Welsh \& Horne (1991): transfer functions are
plotted to the right of the echo image over the same range in time delay, while 
line profiles are plotted below.  

These two cases are different in several respects.  Most strikingly, and rather
deceptively, the line profile in the top plot is extremely broad.  
This occurrence is easily explained by looking at either the echo image 
or the transfer function.  Both show a steep falloff in response, 
which indicates that only the inner high velocity (and thus highly broadened) 
portions of the flow give rise to the line profile.  We call this deceptive
since zooming in on the line profile in the bottom plot would reveal spikes at the same locations
and equally broad emission, but these features are dwarfed by the much higher flux contributed
by the outer portions of the flow.  This flux is in a relatively narrow velocity range around the 
core of the line profile since it originates from lower velocity gas.  

It is useful to draw a comparison with our CM96 benchmark solution (see the Appendix),
which was generated using $A(r) = A_0$.  Notice that its transfer function also has a second
smaller peak like in the top plot of Figure~\ref{fig:disk_case}.
Such can also be seen in the Keplerian disk cases from 
Welsh \& Horne (1991) and P\'erez et al. (1992) and correspond physically 
to the time at which the innermost 
regions of the back side of the accretion flow `come into view'.  
The small dip right before this second peak fills in to become the 
only peak of the transfer function in the bottom plot of Figure~\ref{fig:disk_case}.  
This can be understood by picturing the time delay paraboloid as it sweeps 
toward the back of the disk.  Gas on the sides of disk residing at larger radii than the 
innermost far side gas dominate over either peak due to the $r^2$ weighting. 
 
\subsection{Disk plus disk wind: the role of kinematics} 
\label{sec:kinematics}
We next present calculations for the full PK04 domain 
(that is, from $\theta \approx 8^\circ$ to $\theta = 90^\circ$) in Figure~\ref{fig:diskwind_case}.  
To explore the kinematic effects introduced by the disk wind, in the first case (top plot)
we have again zeroed out the poloidal velocity components and their gradients.
The resulting echo image resembles the disk-only cases in Figure~\ref{fig:disk_case}, although
it is obviously not as wide because the inclination angle has been reduced to $i = 45^\circ$ 
(with the consequence that LoS velocities are smaller).  While a purely rotational wind region
evidently contributes significant emission to line center, the line profile is still prominently 
double-peaked.  

The bottom plot in Figure~\ref{fig:diskwind_case} reveals the primary effect of adding
a poloidal velocity field: there is a marked increase in flux at line center, so that
the line profile is overall single-peaked.  As pointed out by CM96, this effect is 
due to enhanced velocity shear.  Specifically, regions contributing to line center for purely rotational flow 
(namely, gas residing $\phi = 0^\circ$ and $\phi = 180^\circ$) now have higher values of $|dv_l/dl|$
due to the nonzero wind components along the LoS, thereby reducing the optical depth
so that photons can more easily escape.  There are two other noticeable effects: 
increased broadening in the line profile and an overall excess in blue-shifted emission.  
The latter is expected, as we explain in \S{\ref{sec:i_dependence}}.  
The former is again attributable to enhanced velocity shear
because the poloidal velocity field of the wind adds flux to a wide range of LoS velocities.  

While the transfer functions are rather similar in shape, there is overall more response
(by roughly a factor of 3) for that of the bottom plot, which is again an indication of 
significantly enhanced emission due to the poloidal wind components.  
Additionally, the outlines of the echo images are quite similar, the 
main difference being a significant excess in blue-shifted emission for $\tau < 3$ days
in the bottom plot.
(This region is responsible for the blue excess on the line profile; the diagonal feature on the 
blue edge of the echo image at $3 < \tau < 12$ days contributes negligibly.)
Notice, however, the very different shadings of the echo images.  While the purely rotational
case is symmetric with distinct emission patterns, the bottom image is blotchy and lacks any
distinguishing characteristics.

\begin{figure*}                            
 \centering
\includegraphics[width=1.0\textwidth]{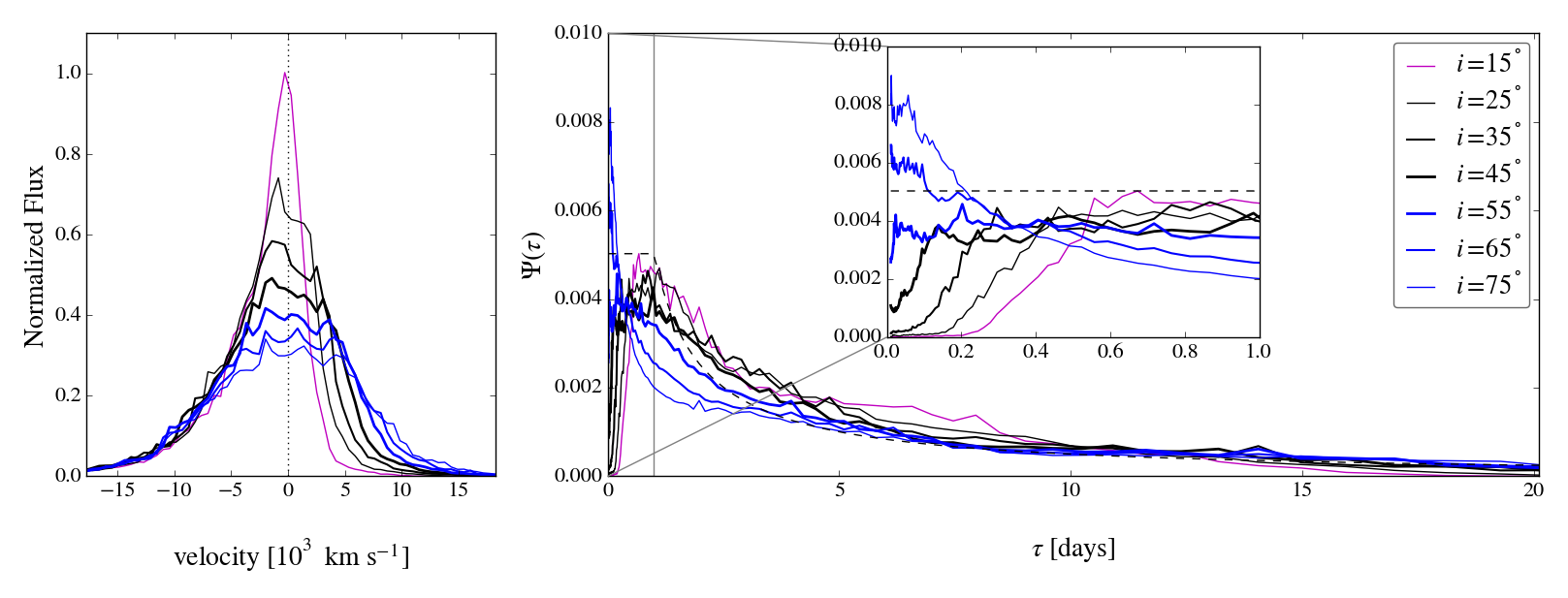} 
\caption{
Parameter study of inclination angle dependence, with the opacity held fixed at $\kappa = 10^4 \kappa_{\rm{es}}$. 
Left panel: line profiles. Right panel: transfer functions.  
The normalization factor $A_0$ is set by normalizing the $i = 15^\circ$ line profile (magenta curve) to unit maximum. 
The jaggedness is due to integrating over a relatively coarse mesh.
The inset plot zooms in on the transfer functions from 0 to 1 day.  
Notice that $\Psi(\tau)$ extends to longer lags as $i$ increases, a property shared with Keplerian disks.  While $\Psi(\tau)$ is sensitive to $i$ for $\tau < 1$ day, this variability would be not be resolved since observational campaigns have cadences of 1 day or more.  
Thus, these transfer functions are effectively degenerate as they all approximately vary as $\tau^{-1}$ beyond 1 day; this scaling is plotted as a dashed line.
The line profiles exhibit a blue-shifted excess at small inclinations, with the red side gradually filling in as $i$ increases and the equatorial wind regions on the far side of the BLR 
become increasingly red-shifted.  Refer to \S{\ref{sec:i_dependence}} for details.
 }
\label{fig:LPTF_istudy}
\end{figure*}
 
\begin{figure*}                            
\includegraphics[width=1.\textwidth]{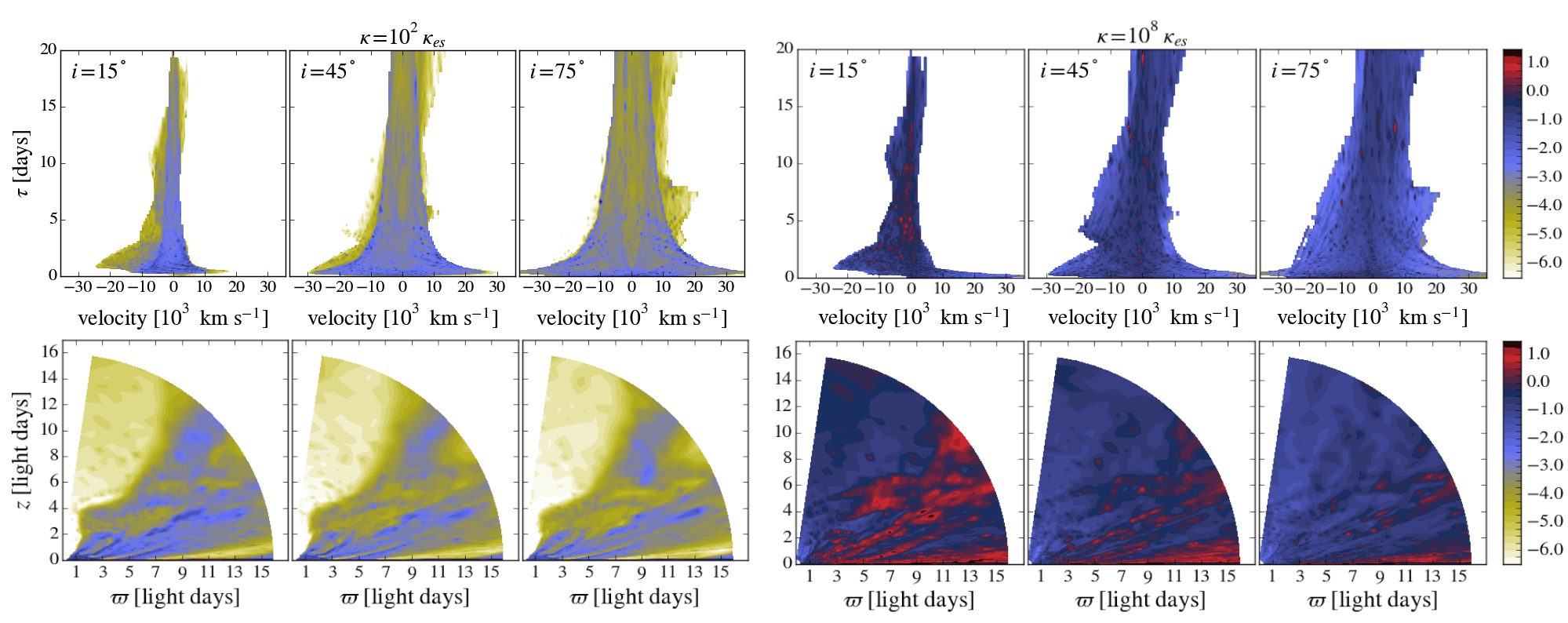} 
\caption{Echo images (top) and corresponding azimuthally averaged volume maps of $\Psi(v_l,\tau)$ 
(bottom) for an optically thin case with $\beta_\nu \approx 1 - \tau_\nu/2$ almost everywhere (left images, with $\kappa = 10^2 \kappa_{\rm{es}}$) and an optically thick
case with $\beta_\nu \approx 1/\tau_\nu$ in the most responsive regions (right images, with $\kappa = 10^8 \kappa_{\rm{es}}$).  
The normalization factor $A_0$ is the same as in Figure \ref{fig:LPTF_istudy}.
The set of images with $\kappa = 10^8 \kappa_{\rm{es}}$ closely resemble the echo image
\emph{sketches} displayed in the bottom row of Figure 1.  Observationally, the outline of the left set of images would
effectively be the bluish virial envelope since the yellow emission is about two orders of magnitude smaller (see the colorbar, which
denotes $\log_{10} \Psi(v_l,\tau)$).  
These plots demonstrate that optically thick and thin lines give rise to qualitatively and quantitatively different echo images.  
 }
\label{fig:volmaps_and_images}
\end{figure*}
 
 \subsection{Dependence on inclination angle} 
\label{sec:i_dependence}
In \S{\ref{sec:sketches}}, we analytically uncovered
an effect of varying the inclination angle:
an excess in blue-shifted emission as $i$ decreases, as would be expected for an outflow.  
To better illustrate this point, in 
Figure~\ref{fig:vlos_maps} we show maps of $v_l$ averaged over each quadrant of $\phi$ for
both a nearly face-on ($i = 15^\circ$) and nearly edge-on ($i = 75^\circ$) viewing angle. 
The poloidal wind field of the PK04 solution is directed nearly radially outward for 
$\theta \gtrsim 70^\circ$, but there is a significant positive $v_z$ component in the region 
$55^\circ \lesssim \theta \lesssim 70^\circ$ (marked by dashed lines).  
This region will therefore appear mostly blue-shifted at low inclinations ($i \lesssim 45^\circ$), 
even on the far side of the BLR, as shown in the left panels.  
This region also has a large velocity shear, so it is very responsive.
Only when it is seen as redshifted on the far side of the disk 
(like in the $i = 75^\circ$ panel) 
can we expect to find line profiles that are roughly symmetric about zero velocity. 
 
These expectations are indeed born out in the line profiles at intermediate inclinations. 
As shown in Figure~\ref{fig:LPTF_istudy}, all the line profiles are single peaked. 
(The jaggedness is due to integrating over a relatively coarse mesh).   
Projection effects tend to reduce $|dv_l/dl|$ at line center
and increase $|dv_l/dl|$ in the line wings as $i$ increases (owing to the dominance
of the rotational component of the velocity), leading to broader and less centrally peaked line
profiles at higher inclinations.  

\begin{figure}                            
 \centering
\includegraphics[width=0.45\textwidth, trim=0 1.5cm 0 1.25cm, clip=true]{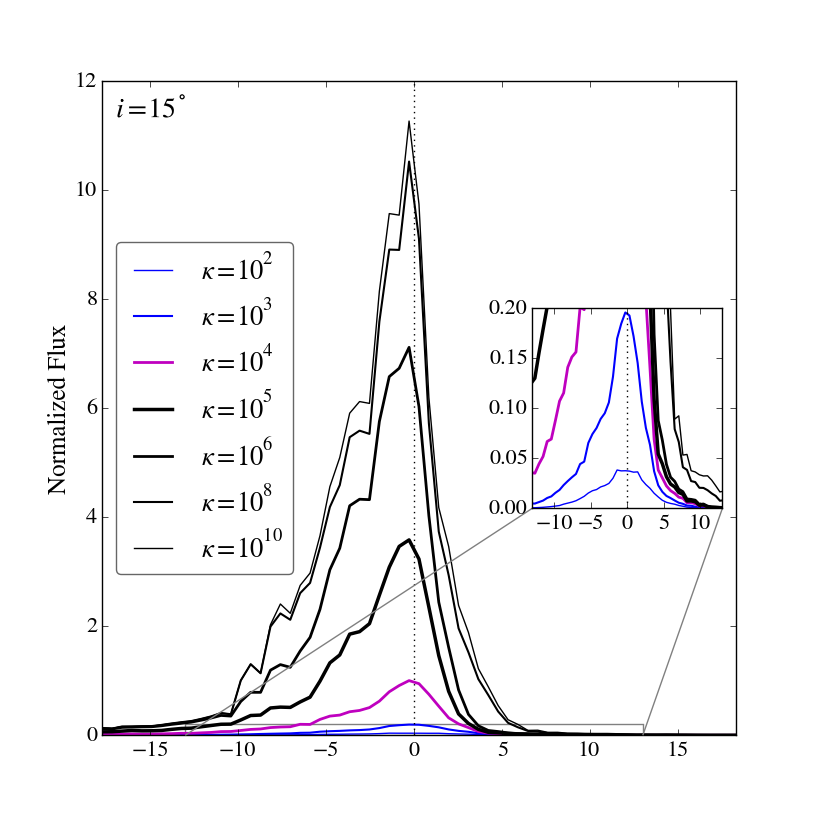}
\includegraphics[width=0.45\textwidth, trim=0 0 0 0.5cm, clip=true]{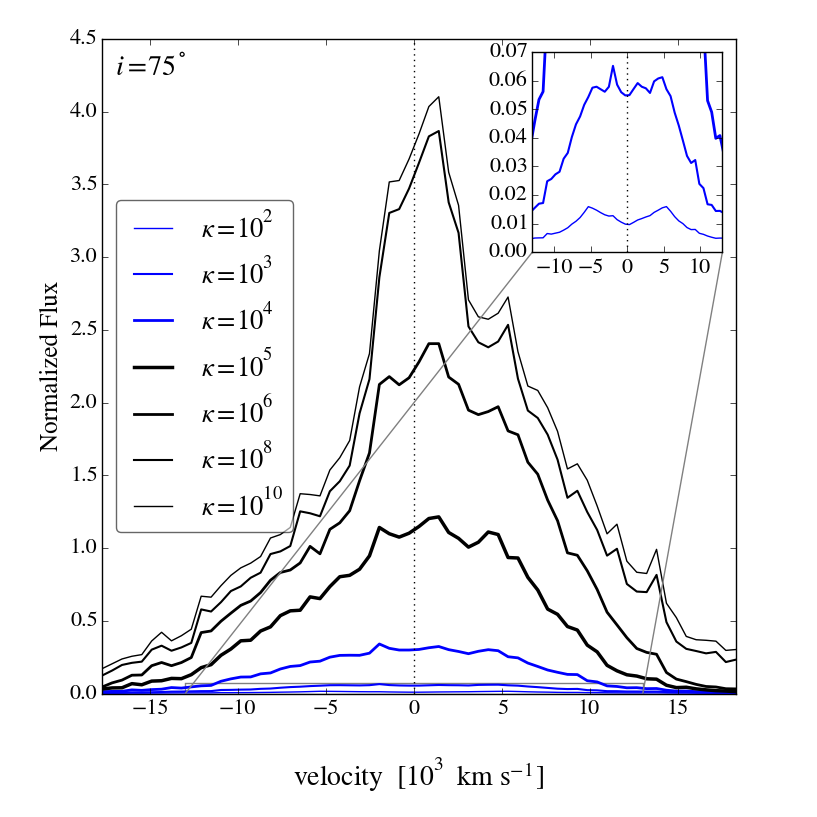}
\caption{
Parameter study of opacity dependence for $i = 15^\circ$ (top) and 
$i = 75^\circ$ (bottom).  The values of $\kappa$ in the
legend are in units of $\kappa_{\rm{es}}$.  
The normalization factor $A_0$ is the same as in Figure \ref{fig:LPTF_istudy}.
The inset plots zoom in on the 
$\kappa = 10^2 \kappa_{\rm{es}}$ and $\kappa = 10^3 \kappa_{\rm{es}}$ line profiles.
When the flow is predominantly optically thin ($\kappa \lesssim 10^4 \kappa_{\rm{es}}$),
the flux increases in proportion to $\kappa$.  For optically thick lines,
the flux is independent of $\kappa$, being set instead by the line of sight 
velocity gradient, $|dv_l/dl|$. 
Refer to \S{\ref{sec:kappa_dependence}} for details.
 }
\label{fig:kappa_study}
\end{figure}

Assessing the dependence of the transfer functions on inclination angle, there is clear property that is shared with a Keplerian disk solution:  as $i$ increases, the transfer function develops an increasingly extended tail.  Recalling Figure~1, this effect is a simple
consequence of the viewing angle, as there is a larger difference in light travel 
times between the front and back sides of the BLR as it is viewed more edge-on.  
Note that our solution domain is about 16.5 light-days across, 
so the maximum time delay is $\sim 33$ light days for 
high inclination angles.  In contrast with those for Keplerian disks, however, the peaks of these transfer functions do not decrease monotonically with increasing $i$ (c.f. Starkey et al. 2015).
A more vertically oriented wind would weaken the response seen at high inclinations for 
$\tau < 1$ day in the inset panel of Figure~\ref{fig:LPTF_istudy}, since in the absence of a strong equatorial wind, only gas in the disk can provide the response. 

\subsection{Dependence on opacity}
\label{sec:kappa_dependence}
To assess the robustness of the results presented thus far, we vary the opacity over 8
orders of magnitude to see how emphasizing and deemphasizing the response of optically
thick and thin regions affects our calculations.  Figure \ref{fig:volmaps_and_images}
shows two sets of echo images for $i = 15^\circ$, $i = 45^\circ$, and $i = 75^\circ$, as well as
azimuthally averaged volume maps of the impulse response function (computed by mapping
$\Psi(v_l,\tau)$ back to $\Psi(r,\theta)$ and then averaging $\Psi(r,\theta)$ over $\phi$ at each $(r,\theta)$).
The left set shows an optically thin case with $\kappa = 10^2 \kappa_{\rm{es}}$, while the right set is an
optically thick case with $\kappa = 10^8 \kappa_{\rm{es}}$.  

These parameter choices sample both of the limiting
regimes discussed in connection with equation \eqref{kappabeta} in \S{\ref{sec:esc_prob}}.  
The integrand for $\Psi(v_l,\tau)$ in the optically thick case is essentially independent 
of the density and opacity, depending only on $|dv_l/dl|$.  Hence, there should be significant dependence
on the inclination angle, with the vertically directed flow region ($55^\circ \lesssim \theta \lesssim 70^\circ$)
becoming more responsive when viewed from lower inclinations.  The azimuthally averaged volume maps
show exactly this.  Meanwhile,
in the optically thin case the integrand of $\Psi(v_l,\tau)$ becomes independent of $|dv_l/dl|$,
varying instead as the product of density and opacity.   
The only quantity that depends on $i$ is the Jacobian, and the left set of volume maps shows that the dependence is relatively weak. 

Referring now to Figure~\ref{fig:kappa_study}, we again find that the line profiles are 
almost always single-peaked.  Only when the gas is predominantly optically thin with 
$\kappa = 10^2 \kappa_{\rm{es}}$, so that the densest equatorial regions dominate over the wind,
does the line profile become double-peaked, as shown in the bottom inset plot.  
(We checked that the line profiles are double-peaked for all $i \gtrsim 20^\circ$ with $\kappa = 10^2 \kappa_{\rm{es}}$.)
Both the line flux and degree of broadening increase monotonically with $\kappa$, and this is true at intermediate 
inclinations as well.  For $\kappa = 10^{2} - 10^4 \, \kappa_{\rm{es}}$, most of the gas is optically
thin $(\beta_\nu \approx 1$), and the impulse response function becomes proportional to 
$k = \kappa \rho \nu_0$.  This explains why the line profiles plotted in blue are so much weaker than the rest.    
For optically thick gas, the flux depends primarily on $|dv_l/dl|$ and the line profiles will show signs of saturation
once $\kappa$ no longer plays a role.  Clearly, the 
$\kappa = 10^{8}$ and $10^{10} \, \kappa_{\rm{es}}$ line profiles
for $i = 15^\circ$ exhibit this saturation, except on the red wings (as shown in the top inset).  
Referring once again to Figure~\ref{fig:vlos_maps}, we indeed find that the red emission originates
from the polar regions, which remain optically thin, explaining the lack of saturation in the red wing.  
On the other hand, the $v_l$-map
for $i = 75^\circ$ shows that the polar regions contribute both red and blue shifted emission.
This is why the $\kappa = 10^{8}$ and $10^{10} \, \kappa_{\rm{es}}$ line profiles for $i = 75^\circ$
show less saturation.  
  
Finally, comparison of these two plots with each other and with 
Figure \ref{fig:volmaps_and_images} provides yet another illustration of the 
net blue shifting effect at low inclinations.  In agreement with 
Figure \ref{fig:volmaps_and_images}, the 
line profiles for $i = 15^\circ$ have higher fluxes overall compared with those for
$i = 75^\circ$ due to there being increased velocity shear in the vertical wind region
when seen from lower inclinations.
The $v_z \cos i$ effect revealed in our analysis of the resonance condition 
causes a blue-shifted excess for $i = 15^\circ$.  For $i = 75^\circ$, the
equatorially concentrated and radially directed wind significantly broadens the
line profile, while the $v_z \cos i$ effect is suppressed on account of $i$ being large.

\subsection{Incorporating photoionization modeling results\\ and accounting for time-dependent effects} %
\label{sec:realistic_models}  
The simple prescription for the responsivity used in this work  
is useful for surveying the properties of a particular BLR model as well as 
for comparing and contrasting different BLR models.  
Upon making a comparison with observations in order to constrain model parameters, 
it will be necessary to calculate the responsivity and opacity
distributions by separately performing photoionization calculations using the properties of 
the BLR model (e.g., temperature and photoionization parameter) as input.   
Although it would not be fully self-consistent, provided the Sobolev approximation applies,
we can then evaluate the impulse response function using equation \eqref{IRFformal}.  
The function $I(r)$ appearing in the integrand becomes, 
\beq I(\mathbf{r}) = \f{1}{4\pi} \pd{j_\nu}{F_X} \f{1 - e^{-\tau_\nu}}{\tau_\nu}, \seq
with the understanding that both $\partial j_\nu/\partial F_X$ and $\kappa$ are 
independently specified as numerical fits or tabulated functions of position.

Both the responsivity and optical depth depend on the density distribution, which may
undergo changes on timescales less than the duration of the observational campaign 
due to the dynamics of individual clumps within the wind.  
In principle, there is no difficulty accounting for time-dependent dynamics by
using a different output from a time-dependent simulation at 
every sampled delay time $\tau$ when constructing the impulse response function
$\Psi(v_l,\tau)$.  Indeed, when computing variable line profiles, this procedure 
should be performed, as comparing results obtained this way with 
those calculated using a single or time-averaged output
can serve as a useful measure of the uncertainty associated with
theoretical line profile predictions.  

Difficulties in accounting for time-dependence arise if the flux variability
inferred from the observed light curve
itself causes significant dynamical changes in the BLR gas, as this violates
the assumption of linearity inherent in equation \eqref{Conv}.  It has recently
been demonstrated using local simulations that the density and acceleration 
of optically thin gas can be appreciably affected by flux variability 
(Waters \& Proga 2016).  If this finding proves true for global calculations as well,
then equation \eqref{Conv} will formally only apply if the flux variability is implicitly 
accounted for in the hydrodynamical simulation.  
In that case, constructing a realistic BLR model will require
solving the equations of radiation hydrodynamics.  

\section{Summary and Discussion}
\label{sec:summary}
In this work, we first concentrated on developing the methodology 
to calculate echo images, line profiles, and transfer functions for 
axisymmetric analytic or simulation-based hydrodynamical models.  
In our calculations, we adopted a simple prescription for the responsivity of
the gas and used a single, representative snapshot of the velocity, density, and
temperature distribution from the PK04 solution.
Our primary goal was to assess the dependence of the observable quantities
on the dominant radiative transfer effects, and we plan to extend this work
and make quantitative comparisons with observations in subsequent papers.  
Our main results are:

(i) Echo images of virialized disk wind solutions overall resemble 
those of randomly oriented cloud models or pure disk models.  
The features introduced by the disk wind are asymmetric and 
dependent on inclination angle and opacity.  However, contributions
from the wind are typically an order of magnitude or more smaller in flux 
outside of the `virial envelope' formed by the rotational component 
(see Figure \ref{fig:volmaps_and_images}).  Because the flux difference
is sensitive to the opacity, optically thick and thin lines can 
be expected to form different echo images.  Wind contributions within
the virial envelope mask any symmetry or distinct emission patterns that are 
characteristic of purely Keplerian motion.  

(ii) Enhanced velocity shear due to adding a wind to a rotationally dominated flow
results in single-peaked variable line profiles.  This result was reported by CM96
and appears to be very robust, at least within the Sobolev approximation.
It was discussed in the context of steady AGN line profiles by \citet{Murray1997} and 
then explored further by \citet{Flohic2012} and \citet{ChajetHall2013}.

(iii) Equatorially confined winds with significant vertical velocity components
are characterized by noticeably blue-shifted echo images and line profiles
for $i \lesssim 45^\circ$.  This tendency is revealed by the resonance condition
itself, i.e. equation \eqref{sketch} is symmetric about $y' = y - (v_z/c) \cos i$
and not $y$, showing that locally the line center frequency is shifted from 
$\nu_0$ to $\nu_0 + \nu_0 (v_z/c) \cos i$.  
The net effect may be observable as a blue-shifted excess on the variable 
(i.e. the rms) line profiles 
after establishing a systemic velocity from the steady line profiles.  
Specifically, we predict that $\rm{H}\beta$ lines are subject to this effect since 
lower ionization lines should originate from this equatorial wind region.  
Higher ionization lines likely originate at smaller radii due to
ionization stratification, or at greater heights above the disk where the 
ionization parameter is larger, which is an instance of vertical stratification
due to a drop off in density with height (see, for example, the discussion 
in Murray \& Chiang 1997; Flohic~et~al.~2012; Giustini~\&~Proga~2012). 

(iv) Transfer functions tend to be degenerate for $\tau > 1$ day.
For low opacities ($\kappa \lesssim 10^4\,\kappa_{\rm{es}}$)
and our fiducial responsivity scaling 
($\partial{j_\nu}/\partial{F_X} \propto \rho\, r^2$), they scale as 
$\Psi(\tau) \propto \tau^{-1}$ for a wide range of inclination angles.  
At higher opacities, the they decline even slower, approximately as 
$\Psi(\tau) \propto \tau^{-1/2}$.  
Distinguishing characteristics are only seen for $\tau \lesssim 1$ day, 
but variability data are not collected on an hourly basis in observational campaigns.  
This suggests that echo images and variable line profiles will be 
the most telling observables
when attempting to extract information about the kinematics of the BLR 
though dynamical modeling.

(v) Despite overall degeneracy with inclination angle and opacity,
transfer functions are quite sensitive to how the responsivity
scales with radius (recall \S{\ref{sec:disk_only}}).  
This is unsurprising, as it has long been asserted that 
transfer functions should prove very useful for constraining the responsivity 
distribution of the BLR (e.g., Goad et al. 1993).  
Our simplified treatment of the responsivity may underestimate
the extended response of emission lines,
considering that the
responsivity parameter $\eta(r)$ (held fixed at 1 in this work), is typically
found to increase from 0 to about 1.2 in detailed photoionization calculations of
both high and low ionization lines (Goad \& Korista 2004; Goad et al. 2012).
Also, transfer functions that decline rather slowly with time 
are inferred in recent observational campaigns
(e.g., Bentz et al. 2010; Grier et al. 2013; Skielboe et al. 2015), 
which moreover reveal that transfer functions can have one or more prominent bumps.  
Such features are expected based on results from photoionization modeling, 
as it has been found that the responsivity can vary sensitively with 
distance and flux state (Korista \& Goad 2004), as well as with the light-curve 
duration (Goad \& Korista 2014).  This brings us to an important point: 
observational campaigns must operate long enough to sample the full range 
of lag durations in the data, as this will be necessary to calibrate the first sets 
of dynamical models that include both photoionization physics and a proper
treatment of hydrodynamics.

\section{Future prospects}
\label{sec:conclusions}
The main success of reverberation mapping to date has been the 
determination of mean time lags, since when combined with the assumption
of virialization, knowing $\langle \tau \rangle$ permits arriving at an estimate for 
the black hole mass via equation \eqref{eq:f}.  Reverberation mapping has
been widely embraced and applied without firmly establishing that the 
responding gas is indeed virialized.  The issue of course is that 
$\langle \tau \rangle$ can be obtained with the \emph{least} demanding application
of reverberation mapping, while validating the virialization assumption  
requires mastery of the \emph{most} demanding type, 
namely echo image reconstruction.  Horne et al. (2004) have called this latter type
high-fidelity reverberation mapping.  

Our results show that 
echo images of the PK04 solution are overall quite similar to other virialized models;
their similarity is due to the dominance of the rotational velocity component in shaping 
the echo images. Therefore, our findings support the notion that observational data must be 
high-fidelity for echo image reconstruction to be of any use in favoring
one virialized model over the next.  

The methodology presented in this work provides 
only the basic framework to make progress on the ultimate 
promise of reverberation mapping, which is to decipher the dynamics of the BLR.  
Looking far ahead, the steps on the theoretical front might 
proceed as follows.  First, high resolution simulations of a BLR model are required 
in order to calculate echo images and variable line profiles to high accuracy.  
Ideally, these simulations
should solve the equations of radiation magnetohydrodynamics and be coupled 
with photoionization calculations in order to best capture
the underlying physics and self-consistently calculate the responsivity.  
Such self-consistency is important as the PK04 solution used in this work
was recently analyzed by \cite{Higginbottom2014} using post-processing 
calculations based on a radiative transfer Monte Carlo - Sobolev code, 
who found that the temperature and ionization parameter in the flow are 
much larger than estimated by PK04.  
The difference comes from the inclusion of
scattered photons from the disk by \cite{Higginbottom2014}, in
contrast to only direct illumination from the central source included
by PK04.  \cite{Higginbottom2014} concluded that the failed wind
cannot be effective in shielding the flow, 
and thus the line-driven wind may turn out to be much weaker unless
other dynamical factors can somehow permit a low ionization fraction
in the wind.

Second, given a fully self-consistent BLR model, the parameters governing this solution
must be varied in order to acquire a set of results with different black hole
masses, accretion rates, gas density distributions, heating and cooling 
prescriptions, ionization networks, etc.
The resulting sets of simulations can then be post-processed one by one,
varying the inclination angle in each case, to finally arrive at a 
large suite of echo images and variable line profiles for this one BLR model.  
For this post-processing step to match the sophistication of the hydrodynamical 
simulations, our current methods will likely not suffice.  
The escape probability formalism may need to be superseded by a full 
Monte-Carlo radiative transfer scheme.  Accounting for extra time delays due 
to multiple scatterings and finite recombination times will require extensions 
to the derivation of the impulse response function, as will accounting for relativistic 
and plasma effects or relaxing the point 
source assumption.  The last of these may be the most serious among the standard 
approximations, considering the recent observations from the NGC 5548 campaigns 
(e.g., \citealt{McHardy2014}; \citealt{Edelson2015}; \citealt{Fausnaugh2015}).

Third and finally, the above process can be repeated for an entirely different BLR model.  
In principle, it is then `just' a matter of comparing each large suite of 
synthetic observables to actual observations in order to narrow down the 
parameter space and see which model performs best using advanced 
fitting techniques such as a Markov Chain Monte Carlo method.  

\section{Acknowledgements}
We thank the referee, K. Korista, for his comments on an early draft of the manuscript,
as well as for a helpful report which improved the final version.
TW and DP thank Sandamali Weerasooriya, George Rhee, and Drew Clausen 
for discussions.  
AK was supported by National Science Foundation through grant AST-1109394.
DP acknowledges support provided by NASA through grants
HST-AR-12835 and HST-AR-12150.01-A from the Space Telescope
Science Institute, which is operated by the Association
of Universities for Research in Astronomy, Inc., under NASA
contract NAS5-26555. 
ME acknowledges support from NSF grant AST-1312686 and thanks the 
Department of Astronomy at the University of Washington, 
where he was based during the early stages of this work, for its hospitality.
Research by A.J.B. is supported in part by NSF grant AST-1412693.

{}

\appendix
\begin{figure}                            
\centering
\includegraphics[width=0.4\textwidth, trim=0 0 0 0, clip=true]{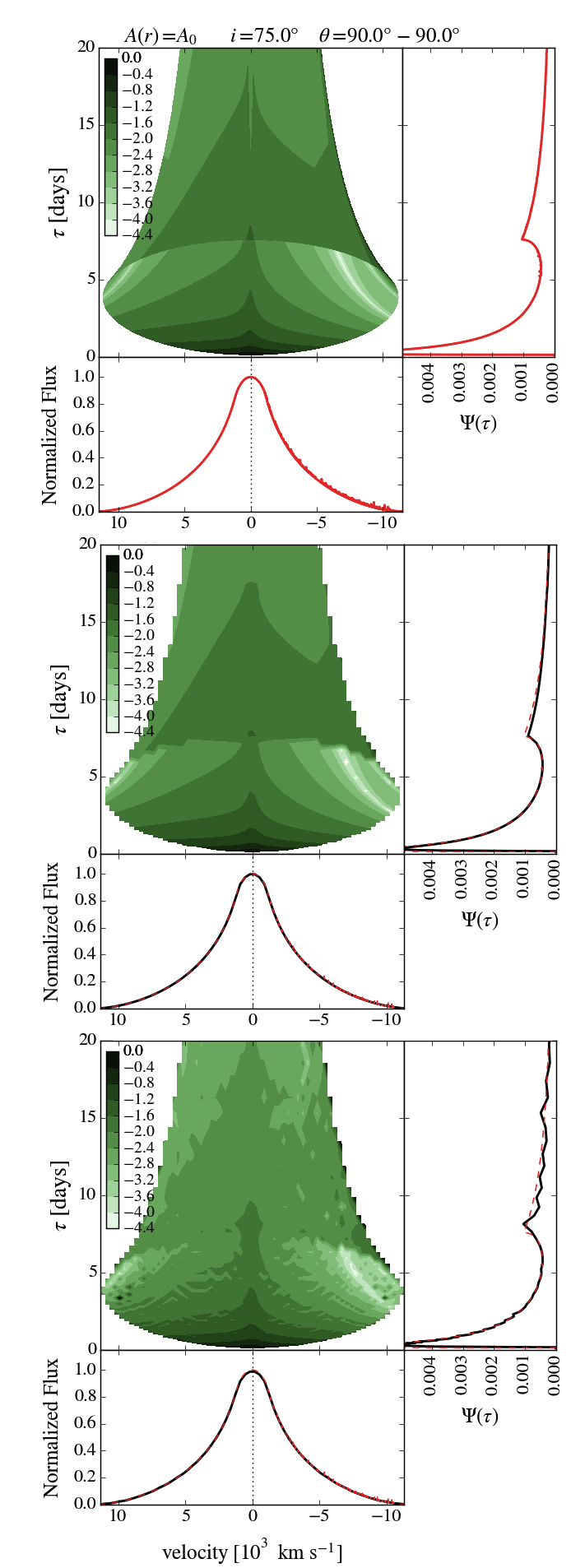}
\caption{
A benchmark calculation using the analytic solution from CM96.  
Top: $[N_\tau,N_{v_l}]=[2048,2048]$ pixel echo image calculated
using the analytic solution.  The line profile (LP) and transfer function (TF) are computed by 
summing over the pixels in the image.  Compare with figs. 2 \& 4 in CM96. 
Center:  $[N_\tau,N_{v_l}]=[128,64]$ pixel echo image calculated using the analytic solution.  
The black solid LP and TF are calculated using our numerical methods on a fine, linearly-spaced 
radial grid with $N_r =4,096$.  
Bottom: $[N_\tau,N_{v_l}]=[128,64]$ pixel echo image calculated using our numerical methods.
The black solid LP and TF are calculated using our numerical methods but using the PK04 
logarithmically-spaced radial grid with $N_r = 100$.  On the center and bottom plots,
the LP and TF from the top plot are over-plotted as dashed red lines.
The normalization factor $A_0$ is set by normalizing the LP in the top plot to unit maximum, 
and colorbars denote $\log_{10} \Psi(v_l,\tau)$.   
Note that CM96 use the opposite sign convention than us, so the blue side is on the right.
 }
\label{fig:CM96}
\end{figure}

 Here we illustrate and benchmark our methods by reproducing the analytic solution
presented by Chiang \& Murray (1996; hereafter CM96).  They considered the case of 
 motion purely in a disk in which $v_\varpi = v_z = 0$, $\theta = \pi/2$, and $\varpi = r$. 
 Hence, equations \eqref{resonance_cdn} read
\begin{align*}
y &= - \f{v_\phi}{c} \sin{\phi} \sin{i} ;\\ 
 t &= \f{r}{c}(1 - \cos{\phi}\sin{i}) .
\end{align*}
Keplerian rotation is assumed, so $v_\phi/c= \sqrt{r_s/2r}$, where $r_s = 2GM_{BH}/c^2$.  
Eliminating $\phi$ between these two equations, 
we find that the resonance condition is cubic in $r$:
\begin{equation*} 
r^3 + \left( \f{r_s \cos^2{i}}{2 y^2} \right) r^2 - \f{r_s c\,t}{y^2} r + \f{r_s (c\,t)^2}{2 y^2} = 0 .
\end{equation*} 
(For $y = 0$, this equation is only quadratic, revealing resonance points $\tilde{r}_\pm = c\,t/(1\mp \sin{i})$.)  
The corresponding values of $\tilde{\phi}$ are those that satisfy \emph{both} the $y$ and $t$ equations above.
Thus, for any desired frequency shift and time delay $(y,t)$, we can algebraically solve for all resonant locations 
$(\tilde{r},\tilde{\phi})$ on the disk.  

It remains to evaluate the LoS velocity gradient $|dv_l/dl|$ and the Jacobian, which by equation \eqref{Jacobian}
also depends on derivatives of the velocity components.  To explore the effects of a wind, CM96 assumed a 
nonzero value for the derivative $dv_r/dr$.\footnote{Note that despite CM96's taking 
$v_r$ to be 0 for all $r$ on the midplane in their eqn. (2), meaning that $dv_r/dr$ is also 0 there, they envisioned a
vertically averaged solution.  Hence, this prescription is consistent with a radial wind region
residing at very small heights above the disk.}  Specifically, it appears they adopted the value 
$dv_r/dr = 3\sqrt{2} v_\phi/r$.  The only other nonzero velocity derivative is 
$dv_\phi/dr= -(v_\phi/2)/r$, giving
\begin{align*}
     \f{dv_l}{dl} &= 3 \f{v_\phi}{r} \sin^2i \cos\phi \left[\sqrt{2}\cos\phi + \f{\sin\phi}{2} \right]; \\
     J &= -\f{v_\phi}{c^2}\sin i \left[ \f{(1 - 3\cos^2\phi)}{2}\sin i + \cos\phi \right].
\end{align*}

We can now evaluate the impulse response function, equation \eqref{integrand}. 
In our formalism, CM96 consider the optically thick limit ($\tau_\nu \gg 1$) and 
$A(r) = A_0$.  
Substituting
$I \rightarrow I\,\delta[\mu - 0]$, we have simply
\begin{equation*}
\Psi(y,t) = \left. \f{I}{|J|} \right|_{(\tilde{r}, \tilde{\phi})},
\end{equation*}
where 
\begin{equation*}
I(\mathbf{r}) = \f{A_0}{4\pi c} \left|\f{dv_l}{dl}\right|.
\end{equation*}
The top plot in Figure \ref{fig:CM96} shows that we have reproduced all of the detailed features of the echo image 
displayed in their Fig. 4, as well as the line profile in their Fig. 2.  

We next solve this problem using our numerical methods.  We discretize the analytically evaluated
velocity components and their derivatives onto the same grid that was used in the PK04 simulation.
In velocity-delay space, the PK04 grid spans a width of $[-36, 36] \times 10^3\:\rm{km\,s^{-1}}$ and 
a height of 33 days, while the CM96 solution spans a width of $[-11.5, 11.5] \times 10^3\:\rm{km\,s^{-1}}$ 
and a height of 780 days.  For the PK04 solution, we found the optimal image resolution to be $128 \times 128$ 
pixels spaced linearly in velocity (i.e. each pixel spans $0.56 \times 10^3\:\rm{km\,s^{-1}}$) and logarithmically in 
time delay.  To make a fair comparison, in the center and right plots of Figure \ref{fig:CM96} we use the same time 
delay resolution (128 pixels covering 33 days), but we use just half the resolution (i.e. 64 pixels) to cover CM96's 
smaller velocity range.  Analytically evaluating $\Psi(y,t)$ on this grid gives the result shown in the center plot of 
Figure \ref{fig:CM96}, while numerically evaluating $\Psi(y,t)$ yields the bottom plot.  The interpolation procedure 
tends to blur the image patterns somewhat, while for $\tau \gtrsim 7$ days there is also a small reduction in 
brightness that is likely more due to the logarithmic PK04 grid.

The transfer functions and line profiles plotted in red on the top plot serve as our reference solutions and are 
calculated using the indirect method (recall \S{\ref{sec:LPTFcalc}}), in which we simply sum over the image 
using equation \eqref{trap_rule}.  We employ our direct integration method to calculate the line profiles and 
transfer functions plotted in black on the center and bottom plots (and we overplot the reference solutions 
as red dashed lines).  For the center plot, we use a fine linearly spaced grid to carry out the numerical 
integration over radius, while we use the much coarser but logarithmic PK04 grid for the bottom plot.
Notice that the logarithmic spacing causes some numerical noise on the transfer function
beyond the second peak, demonstrating that the grid, as opposed to the PK04 solution, is to blame for 
much of the jaggedness in Figure~\ref{fig:LPTF_istudy}. 

We further benchmarked our code against a spherically symmetric wind model from
Welsh \& Horne (1991).  This test was needed to verify our integration over $\mu$
since the CM96 solution does not test this aspect of our code.  We again found
an exact match at high resolutions, and the echo image, line profile, and transfer function
were all sufficiently reproduced upon using the PK04 grid.  We conclude from these tests
that high resolution simulations will be needed when there are steep gradients in the 
velocity or density fields in order to obtain smooth line profiles and transfer functions.  

\end{document}